\documentclass[a4paper,11pt]{article}
\usepackage{jheppub}
\usepackage{mathrsfs}
\usepackage{amsfonts}
\usepackage{setspace}
\usepackage{cellspace}
\usepackage{amsmath,bm}
\usepackage[colorlinks=true,linkcolor=blue]{hyperref}
\usepackage{xcolor}
\usepackage{epsfig}
\usepackage{slashed}
\usepackage{caption}
\usepackage{hhline,multirow,tabularx}  
\usepackage{dcolumn}    
\usepackage{url}        
\usepackage{braket}

\title{Asymptotics of spin-spin correlators weighted by fermion number measurements with low rapidity threshold in the 2D Ising free-fermion QFT}
\author[a]{Yizhuang Liu}

\affiliation[a]{Institute of Theoretical Physics,
Jagiellonian University, 30-348 Kraków, Poland}

\emailAdd{yizhuang.liu@uj.edu.pl}

\abstract {In the work, we study the averaged number of massive fermions above a low rapidity threshold $Y$, underlying the form-factor expansions of the spin-spin two-point correlators at an Euclidean distance $r$, in the 2D Ising QFT at the free massive fermion point. Despite the on-shell freeness, the spin operators are still far away from being Gaussian, and create particles in the asymptotic states with complicated correlations. We show how the number observables can still be incorporated into the integrable Sinh-Gordon/Painleve-III framework and controlled by linear differential equations with two variables $(r,Y)$. We show how the differential equations and the information of two crucial {\it scaling functions} arising in the $r\rightarrow 0$, $e^{Y}r={\cal O}(1)$  {\it scaling limit}, can be combined to fully determine the small-$r$ asymptotics of the observables, in the {\it $\lambda$-extended} form. The scaling functions, on the other hand, are obtained by summing the exponential form-factor expansions directly, generalizing the traditional Ising connecting computations. We show carefully, how the singularities cancel in the physical value limit $\lambda \pi \rightarrow 1$ and how the power-corrections that collapse at this value can be resummed. In particular, we show for the physical $\lambda$-value, the scaling functions are related to  integrated four-point functions in the Ising CFT and continue to control the asymptotics of the number-observables in the scaling limit up to ${\cal O}(r^3)$. }
\date{\today}

\begin{document}
\maketitle
\flushbottom
\section{Introduction}
The exact analysis of the spin-spin two-point correlators in the 2D Ising model~\cite{PhysRev.149.380,PhysRev.155.438,PhysRev.164.719, Wu:1975mw} is among one of the most beautiful works in 20th-century theoretical physics. One starts with a simple lattice action, and ends up with an Euclidean correlation function in a massive quantum field theory~\cite{McCoy:1977er}, that has spectacular power-and-logarithm asymptotics at small distance. We now know how such asymptotics could be explained: the largest algebraic term $r^{-\frac{1}{4}}$ belongs to the Ising-CFT,  the power corrections generated by the CFT perturbation theory with operator condensates~\cite{Shifman:1978bx,Shifman:1978by,David:1983gz,Novikov:1984rf,Zamolodchikov:1990bk,Lukyanov:1996jj}, and the IR/UV ambiguities of the coefficient functions and the condensates reflect through the wonderful logarithms. In fact, this picture of short distance asymptotics of Euclidean correlators is widely believed to hold in a larger class of massive quantum field theories with UV fixed points, including the QCD~\cite{Shifman:1978bx,Shifman:1978by,Novikov:1984ac}. 

On the other hand, it is still not clear now, whether the notion of UV fixed points and the related perturbative frameworks could apply to a broader class of high-energy limits, such as the notorious Regge limit of forward scatterings, a limit that might not appear to resemble the Euclidean short-distance limit at all. It is natural to ask the following questions: is the QCD Regge limit described or even related to the perturbative QCD? If yes, how to write out the Regge asymptotics explicitly? If not, what are the alternatives that allow the precise computation of such asymptotics, if they ever exist? These questions still have no clear answers, after the 50-year discovery of the asymptotically freedom.

Of course, the Euclidean two point functions and the Regge limit are two extreme examples to show the intricate nature of  the high-energy limit in the broad sense. In between, there are many quantities that are more complicated than an Euclidean two-point function, while still simpler than the Regge limit. One type of such quantities are certain non-inclusive measurements underlying the $e^+e^-\rightarrow X$ type inclusive processes. If summed over all final states, one ends up again at a two-point function and the large $Q^2$ injected by the operator indicates one has perturbative asymptotics. What if one ``opens up'' the two point function by introducing some non-trivial measurements on the asymptotic states?  There are hopes (see~\cite{Moult:2025nhu} for a recent review) that for certain IR-safe measurements, in the large $Q^2$ limit, one continues to have perturbative aymptotics that can be computed in the short distance massless formulations, despite that the massless particles in such formulations are in no way, the real asymptotic states in the measurements. Remember, it is the exponential asymptotics of correlation functions that identify the stable particles. From the perspective of the underlying massive theories, the convergences of real measurements to the massless limits are therefore extremely non-trivial {\it connecting problems}, connecting the UV and IR descriptions of the same theory. As such, it would be helpful to find explicit examples in integrable 2D QFTs for which such connecting problems could be worked out. 

There are many integrable 2D QFTs in which the form factor expansions~\cite{Smirnov:1992vz,Babujian:2006km} of the two-point correlators can be written down explicitly up to an arbitrary finite particle number (for example~\cite{Balog:1992jq,Balog:1994np,Babujian:2013roa}), at least in principle. But for most of such theories, the connecting problem for correlators remains intractable, except in the large $N$ expansions~\cite{Campostrini:1991eb,Campostrini:1993gya,Beneke:1998eq,Marino:2024uco,Liu:2025bqq}. The only non-large-$N$ exceptions are the Sine-Gordon QFT/2D Ising QFT at the free-fermion point~\cite{McCoy:1976cd,Tracy:1990tn}. It happens that the underlying lattice modes can also be made free fermionic and posses large amounts of conserved charges or symmetries~\cite{Onsager:1943jn}, supporting closed relations already at the level of lattice correlators~\cite{PERK19803,McCoy:1980as,10.3792/pjaa.56.405,PERK19841}. On the other hand, the on-shell freeness also allows to simplify the form factor expansions of local correlators into Fredholm-determinants, enabling the derivation of integrable differential equations in the continuum  limit~\cite{McCoy:1976cd,Bernard:1994re,Babelon:1992sn}, obtainable also from the lattice level identities~\cite{Au-Yang2002}.

It turns out that such differential equations are rather universal. For example, the derivation given in~\cite{Tracy_1996} for the Sinh-Gordon relies only on the 
$(x+y)^{-1}$ form of the non-factorizable part of the kernel function. As a result, such equations are usually unable to fully fix the small distance expansions, but the high-order terms were allowed to express through the first few undetermined coefficients, which we call the connecting coefficients. The very explicit expressions with many details in the form factor integrals can then be utilized to compute them. The computations of the Ising connecting coefficients following this route were performed and presented in details in~\cite{McCoy:1976cd,Tracy:1990tn}. They heavily rely on the property of an integral operator and the related Mellin-Barnes representations.  In particular, the logarithms in the short distance expansion are reflected by the singularities of the connecting coefficients and their cancellation with the ``power corrections'', at the physical value of the $\lambda$-extension parameter keeping track of the particle numbers.

The Euclidean two-point functions could be regarded as mimicking the inclusive total cross sections. Indeed, the form-factors squared are summed in a very inclusive manner under a very smooth weight $e^{-Er}$, introducing the energy cutoff through the inverse distance $r^{-1}$ that plays a role similar to the total energy $Q$. As such, one can try to introduce non-trivial measurements on the final states and study the short distance limits of the weighted two-point functions. The purpose of this work is to provide an explicit example of a non-trivial one-particle measurement, for which the connecting problem can be worked out explicitly in the 2D Ising free-fermion QFT, generalizing the Ising connecting computations of the inclusive two-point function.

More precisely, we will compute the averaged number of particles whose rapidities are above a given lower threshold $Y$, in the form factor expansion of the Euclidean spin-spin correlators. In the rest of the paper, we call these averages the ``number observables'' or simply ``observables''. We are interested in the small-$r$ limit at a fixed $Y$, as well as the $r\rightarrow0$, $re^{Y}={\cal O}(1)$ {\it scaling limit}. We show in detail how such measurement can be incorporated into the Sinh-Gordon framework, at the price of adding the $Y$ as a new variable in the differential equations. Due to the two-variable nature, the connecting coefficients require more information to fully determine the small-$r$ asymptotics. We will see that two {\it scaling functions}, encoding the largest contributions in the scaling limit for the $\lambda$-extended form factor expansions when $0<\lambda\pi<1$ , exactly play the role of the connecting coefficients. As the original Ising connecting problem, they are obtained by summing the form factor expansions using the knowledge of an integral operator. We will show how the small-$r$, fixed $Y$ asymptotics and the small-$r$, fixed $e^{Y}r$ asymptotics can be determined given the knowledge of the scaling functions, and how the singularities cancel in the physical value limit $\pi\lambda \rightarrow 1$, generating the logarithms in the expansion. In particular, we show for the physical $\lambda$ value, the scaling functions remain finite and can be extracted from certain Ising-CFT four-point correlators. They dominate the asymptotics of the number observables in the scaling limit up to ${\cal O}(r^3)$, at the physical $\lambda$ value. 

The organization of the paper is as follows. In Sec.~\ref{sec:def}, we review the Sinh-Gordon differential equations for certain Fredholm determinants and introduce our observables in details. We present their exponential form factor expansions and establish the $(r,Y)$ two-variable differential equations. For convenience of the readers, after defining our observables, at the end of the section we summarize the main results of this work in the subsection.~\ref{sec:summarymain}. The Sec.~\ref{sec:analysis} is the bulk of the work. We first review the short distance asymptotics of the inclusive two point function, emphasizing the $\lambda\pi \rightarrow1$ singularities and their cancellation. We then introduce the scaling limits and the scaling functions in Sec.~\ref{sec:scaling}. We use the knowledge of an integral operator to sum the form factor expansions in the scaling limit and obtained the two-parameter Mellin-Barnes representations for the scaling functions at a generic $\lambda$. The Mellin-Barnes, providing crucial boundary conditions for the $(r,Y)$ two variable differential equations, will be used to generate systematically the small-$re^{Y}$ expansion of the scaling functions. In  Sec.~\ref{sec:scaling2}, we further simplify the scaling functions at the physical $\lambda$ parameter and reduce the number of Mellin-Barnes variables to one~\footnote{Actually, after finishing the first version of the work, we noticed that they can be further simplified into the Bessel $K$ functions.}. The one-variable Barnes representations then allow to establish the connection to an integrated four-point-function in the Ising CFT.  Given the knowledge of the scaling functions, we then perform the small-$r$ expansions at a fixed $Y$ in Sec.~\ref{sec:smallr}, and at a fixed $t=\frac{r}{2}e^Y$ in Sec.~\ref{sec:scalinglimit}. We show how the ``power-corrections'' that collapse in the $\lambda\pi \rightarrow 1$ limit can be resummed using the differential equations, and how the singularities cancel among the various contributions, leading to finite results with logarithms at each power in the small $r$ expansion at $\lambda\pi=1$. The asymptotics of the observables in the scaling limit are given by the Eq.~(\ref{eq:finalasymptotics}) up to ${\cal O}(r^3)$.  Finally, we conclude and make further comments in Sec.~\ref{sec:conclude}. In the appendix.~\ref{sec:cylinder}, we compute the number observables in the Ising-CFT on a cylinder and show how the scaling functions $\mathfrak{f}_{\pm}(t)$ appear~\footnote{The author thanks the anonymous referee for suggesting the connection between $\mathfrak{f}_-(t)$ and CFT correlators with Ramond fermions}. The other two appendices collect useful technical details.

\section{Fredholm determinants and the fermion number observables}\label{sec:def}
To start the presentation, we first review the mKdV/Sinh-Gordon hierarchy~\cite{Bernard:1994re,Babelon:1992sn,Tracy_1996} for certain Fredholm determinants. We introduce the following operator on $L^2(\mathbb{R}_+)$
\begin{align}\label{eq:defKgeneral}
K(x,y)=\frac{E(x)E(y)e^{-\frac{1}{2}\left(zx+\bar zx^{-1}\right)-\frac{1}{2}\left(zy+\bar zy^{-1}\right)}}{x+y} \ .
\end{align}
Here, for simplicity we require $E(x)$ to be a bounded piecewisely smooth function whose precise form is flexible, while ${\rm Re}(z),{\rm Re}(\bar z)>0$. We introduce the following potential functions for $0<\lambda<\frac{1}{\pi {\rm sup}_{x>0}|E|}$
\begin{align}
&\phi_{\pm}(z,\bar z,\lambda)={\rm Tr} \ln (1\pm \lambda K)  \ , \\
&\varphi(z,\bar z,\lambda)=\phi_+(z,\bar z,\lambda)-\phi_-(z,\bar z,\lambda) \ , \\ 
&\chi(z,\bar z,\lambda)=\phi_+(z,\bar z,\lambda)+\phi_-(z,\bar z,\lambda) \ , 
\end{align}
as well as the determinants 
\begin{align}
F_{\pm}(z,\bar z,\lambda)=\frac{1}{2}\bigg({\rm det}(1+\lambda K) \pm {\rm det}(1-\lambda K)\bigg)=\frac{1}{2}(e^{\phi_+} \pm e^{\phi_-})  \ .
\end{align}
The derivatives are introduced as $\partial=\frac{\partial}{\partial z}$ and $\bar \partial=\frac{\partial}{\partial \bar z}$. Then, it has been shown~\cite{Tracy_1996} that the potential functions satisfy the following Sinh-Gordon equations
\begin{align}
&\frac{\partial^2}{\partial z\partial \bar z}\varphi=\frac{1}{2}\sinh 2\varphi \ , \label{eq:SG1}\\ 
&\frac{\partial^2}{\partial z\partial \bar z}\chi=\frac{1}{2}\left(1-\cosh2\varphi\right) \ , \label{eq:SG2}
\end{align}
as well as the auxiliary equations
\begin{align}
&\partial^2\chi=(\partial \varphi)^2 \ ,  \label{eq:auxi1}\\
&\bar \partial^2\chi=(\bar \partial \varphi)^2 \ . \label{eq:auxi2}
\end{align}
They are only the first few among a set of relations called the integrated mKdV hierarchy, connecting derivatives of the potential functions when more parameter are added into the kernel, but in this work we only need the parameters $z$ and $\bar z$, corresponding to $-t_{\pm1}$ in~\cite{Tracy_1996}.  Notice that the above are very general properties of the operators Eq.~(\ref{eq:defKgeneral}) and are not sensitive to the precise form of the $E(x)$, as far as $\lambda{\rm Tr}(K)<1$. 

On the other hand, it is the simplest choice of $E(x)$, $E(x)=1$ that leads to the spin-spin correlation functions of the 2D Ising model in the $T\rightarrow T_c^{\mp}$ massive scaling limit at zero magnetic field~\cite{Wu:1975mw,McCoy:1976cd,PALMER1981329}. In particular, if we expand the Fredhom determinants using the explicit product formula for the Cauchy-determinants, then 
\begin{align} \label{eq:Fplus}
F_{+}(z,\bar z,\lambda)=&1+\sum_{k=1}^{\infty}\frac{\lambda^{2k}}{(2k)!}\int_0^{\infty} dx_1...dx_{2k} {\rm det}_{2k\times 2k}\frac{1}{x_i+x_j}e^{-\frac{r}{2}\sum_{i=1}^{2k}(x_i+x_i^{-1})} \nonumber \\ 
=&1+\sum_{k=1}^{\infty}\frac{\lambda^{2k}}{2^{2k}(2k)!}\int_0^{\infty} \prod_{i=1}^{2k}\frac{dx_i}{x_i}\prod_{i<j}\left(\frac{x_i-x_j}{x_i+x_j}\right)^2 e^{-\frac{r}{2}\sum_{i=1}^{2k}(x_i+x_i^{-1})} \ , 
\end{align}
and similarly 
\begin{align} \label{eq:Fminus}
F_{-}(z,\bar z,\lambda)=\sum_{k=0}^{\infty}\frac{\lambda^{2k+1}}{2^{2k+1}(2k+1)!}\int_0^{\infty} \prod_{i=1}^{2k+1}\frac{dx_i}{x_i}\prod_{i<j}\left(\frac{x_i-x_j}{x_i+x_j}\right)^2 e^{-\frac{r}{2}\sum_{i=1}^{2k+1}(x_i+x_i^{-1})} \ .
\end{align}
Here, we have introduced the rescaled distance $r^2=4z\bar z$, which is the distance measured in the unit of the free fermion mass $m=1$. These representations for $F_{\pm}$ are then the {\it form factor expansions} for the $T\rightarrow T_c^{\mp}$ rescaled spin-spin correlators, when $\lambda=\frac{1}{\pi}$. Indeed, if we change to the rapidity variables $x_i=e^{\theta_i}$, then we have the following form factors~\cite{cmp/1103904079,Berg:1978sw}
\begin{align}
&\bigg|\langle \theta_1...\theta_{2n}|\sigma_{T_c^{-}}(0)|\Omega\rangle\bigg|^2=\prod_{i<j}\left(\frac{e^{\theta_i}-e^{\theta_j}}{e^{\theta_i}+e^{\theta_j}}\right)^2=\prod_{i<j} \tanh^2 \frac{\theta_i-\theta_j}{2} \ , \\
&\bigg|\langle \theta_1...\theta_{2n+1}|\sigma_{T_c^{+}}(0)|\Omega\rangle\bigg|^2=\prod_{i<j}\left(\frac{e^{\theta_i}-e^{\theta_j}}{e^{\theta_i}+e^{\theta_j}}\right)^2=\prod_{i<j} \tanh^2 \frac{\theta_i-\theta_j}{2} \ .
\end{align}
In terms of them, Eq.~(\ref{eq:Fplus}) and Eq.~(\ref{eq:Fminus}) can be written as 
\begin{align}
&F_{+}\left(r,\lambda=\frac{1}{\pi}\right)=1+\sum_{n\in 2\mathbb{Z}_{>0}}^{\infty}\frac{1}{n!}\int_{-\infty}^{\infty}\prod_{i=1}^{n}\frac{d\theta_i}{2\pi} \bigg|\langle\theta_1..\theta_{n}|\sigma_{T_c^{-}}(0)|\Omega\rangle\bigg|^2e^{-r\sum_{i=1}^{n}\cosh \theta_i} \ , \\
&F_{-}\left(r,\lambda=\frac{1}{\pi}\right)=\sum_{n\in 2\mathbb{Z}_{\ge 0}+1}^{\infty}\frac{1}{n!}\int_{-\infty}^{\infty}\prod_{i=1}^{n}\frac{d\theta_i}{2\pi} \bigg|\langle\theta_1..\theta_{n}|\sigma_{T_c^{+}}(0)|\Omega\rangle\bigg|^2e^{-r\sum_{i=1}^{n}\cosh \theta_i} \ .
\end{align}
From the above, the interpretation of $F_{\pm}$ as the spectral decomposition of Euclidean correlators become manifest. In particular, the spin operator $\sigma_{T_c^+}$ creates asymptotic states with odd number of massive fermions, while the disorder operator $\sigma_{T_c^-}$ creates even numbers. The massive fermions are free and real, with the normalization $\langle \theta|\theta' \rangle=2\pi\delta(\theta-\theta')$. On the other hand, the spin operators are far from being Gaussian and create particles in the asymptotic states in a sufficiently complicated manner, with non-trivial correlations among the created particles. 

\subsection{The fermion number observables}
It is clear that the two-point function themselves can be regarded as inclusive measurements on the the created particles, in the sense that all possible final states are summed over under the weight $e^{-Er}$ for the Euclidean correlator. As such, they can be regarded as the coordinate space version of the $e^{+}e^{-} \rightarrow X$ type inclusive total cross sections where the $r^{-2}$ plays the role of the total energy $Q^2$.  Clearly, like the real process, one can introduce more complicate measurements on the final state particles. In this work we will investigate a specific measurement, which counts the number of particles above a given low rapidity threshold.  Namely, in the form factor expansion, we introduce an additional weighting function
\begin{align}
O(n,Y,\theta_i)=\sum_{i=1}^n \theta(\theta_i-Y) \ , 
\end{align}
where $\theta(x)=0, x\le 0; \  \theta(x)=1, x>0$ is the standard Heaviside $\theta$ function,  in the phase-space integrals over the $n$-particle rapidities
\begin{align}
\int_{-\infty}^{\infty}\prod_{i=1}^{n}\frac{d\theta_i}{2\pi} \rightarrow \int_{-\infty}^{\infty}\prod_{i=1}^{n}\frac{d\theta_i}{2\pi}\sum_{i=1}^n \theta(\theta_i-Y) \ ,
\end{align}
and normalize the weighted two point functions to the un-weighted ones. Clearly, this can be regarded as the averaged number of particles whose right-moving momentum are greater than $\sinh Y$. For $Y=0$, this is just half of the averaged number of total particles. 

Here we show how the weighted sum can be incorporated into the Sinh-Gordon formalism. The trick is to use the following function in the operator 
\begin{align}
E_{\alpha}(x)=\bigg(1+\alpha \theta(x-e^Y)\bigg)^{\frac{1}{2}} \  . 
\end{align}
Here $|\alpha|<1$ is a real number. We denote the Fredholm operator defined according to Eq.~(\ref{eq:defKgeneral}) using the $E(x)=E_{\alpha}(x)$ as $K_{\alpha}$. After introducing the function $E_{\alpha}$, the form factor expansions for $F_{\pm}$ defined with $K_{\alpha}$ are then modified to 
\begin{align}
\int_{-\infty}^{\infty}\prod_{i=1}^{n}\frac{d\theta_i}{2\pi} \rightarrow \int_{-\infty}^{\infty}\prod_{i=1}^{n}\frac{d\theta_i}{2\pi}\prod_{i=1}^n \left(1+\alpha \theta(\theta_i-Y)\right) \ .
\end{align}
We then take derivative with respect to $\alpha$ at $\alpha=0$, this leads exactly to the desired measurement
\begin{align}
\frac{d}{d\alpha}\prod_{i=1}^n \left(1+\alpha \theta(\theta_i-Y)\right) \bigg|_{\alpha=0}=\sum_{i=1}^n \theta(\theta_i-Y) \ . 
\end{align}
On the other hand, for any $\alpha$, the Sinh-Gordon hierarchy remains to be valid. Furthermore, after rescaling the $x_i$, we can transmute the $Y$ dependency to $\bar z$ and $z$:
\begin{align}
E_{\alpha}(x) \rightarrow \bigg(1+\alpha \theta(x-1)\bigg)^{\frac{1}{2}} \ , \ z\rightarrow z e^{Y}, \  \bar z \rightarrow \bar ze^{-Y} \ . \label{eq:boost}
\end{align}
As such, without losing of generality we start with $\bar z=z=\frac{r}{2}$, and express the $z$, $\bar z$ derivatives in terms of $r$ and $Y$. Denoting the $\alpha$ dependent potential functions defined with $K_{\alpha}$ as $\varphi(r,Y,\alpha)$, $\chi(r,Y,\alpha)$, we have 
\begin{align}
\bigg(\frac{\partial^2}{\partial r^2}+\frac{1}{r}\frac{\partial}{\partial r}-\frac{1}{r^2}\frac{\partial^2}{\partial Y^2}\bigg)\varphi(r,Y,\alpha,\lambda)=\frac{1}{2}\sinh 2\varphi(r,Y,\alpha,\lambda) \ , \\ 
\bigg(\frac{\partial^2}{\partial r^2}+\frac{1}{r}\frac{\partial}{\partial r}-\frac{1}{r^2}\frac{\partial^2}{\partial Y^2}\bigg)\chi(r,Y,\alpha,\lambda)=\frac{1}{2}\left(1-\cosh 2\varphi(r,Y,\alpha,\lambda)\right) \ .
\end{align}
Now, by taking the $\alpha$ derivative and using $\varphi(r,Y,\alpha=0,\lambda)=\varphi(r,\lambda)$, $\chi(r,Y,\alpha=0,\lambda)=\chi(r,\lambda)$, we obtains the linear differential equations for
\begin{align}
{\cal Q}(r,Y,\lambda)=\frac{d}{d\alpha}\varphi(r,Y,\alpha,\lambda)\bigg|_{\alpha=0} \ , \\
{\cal H}(r,Y,\lambda)=\frac{d}{d\alpha}\chi(r,Y,\alpha,\lambda)\bigg|_{\alpha=0} \ ,
\end{align}
as
\begin{align}
&\bigg(\frac{\partial^2}{\partial r^2}+\frac{1}{r}\frac{\partial}{\partial r}-\frac{1}{r^2}\frac{\partial^2}{\partial Y^2}\bigg){\cal Q}(r,Y,\lambda)=\cosh 2\varphi(r,\lambda){\cal Q}(r,Y,\lambda) \ , \label{eq:lineardifQ} \\ 
&\bigg(\frac{\partial^2}{\partial r^2}+\frac{1}{r}\frac{\partial}{\partial r}-\frac{1}{r^2}\frac{\partial^2}{\partial Y^2}\bigg){\cal H}(r,Y,\lambda)=-\sinh 2\varphi(r,\lambda){\cal Q}(r,Y,\lambda) \label{eq:lineardifH} \ .  
\end{align}
It is also convenient to introduce the function $\eta(r,\lambda)=e^{-\varphi(r,\lambda)}$, which satisfies the Painleve equation of the third kind~\cite{McCoy:1976cd}. With out losing of generality, from now on we omit the $\lambda$ dependencies in arguments of the functions, with the understanding that they all depend on $\lambda$. In terms of the above, the observables can be expressed as
\begin{align}
&N_{\pm}(r,Y) \equiv \frac{1}{F_{\pm}(r)}\frac{d}{d\alpha} F_{\pm}(r,Y,\alpha)\bigg|_{\alpha=0}=\frac{{\cal H}(r,Y)+{\cal Q}(r,Y)}{2}\mp\frac{\eta(r)}{1\pm\eta(r)}{\cal Q}(r,Y) \ . \label{eq:observable}
\end{align}
In the rest of the paper, we investigate the small $r$ and large $Y$ expansions of the ${\cal H}$ and ${\cal Q}$ functions in order to obtain the observables $N_{\pm}$. 

Before proceeding further, let's present the exponential form factor expansions for the potential functions and their $\alpha$ derivatives. It turns out that the exponential form factor expansions are easier to dealt with in the small $r$ limit. Using the expansion of the potential functions $\phi_{\pm}(r,Y,\alpha)={\rm Tr}\ln(1\pm \lambda K_{\alpha})$, it is straightforward to show that
\begin{align} \label{eq:expformphipm}
\phi_{\pm}(r,Y,\alpha)=-\sum_{n=1}^{\infty}\frac{(\mp \lambda)^n}{n}\int_0^{\infty}\prod_{i=1}^n \left(1+\alpha \theta(x_i-e^{Y})\right)dx_i \prod_{j=1}^n\frac{e^{-\frac{r}{2}\left(x_j+x_j^{-1}\right)}}{x_j+x_{j+1}} \ .
\end{align}
As such, after introducing the $\alpha$-derivatives
\begin{align} \label{eq:expformY}
\Phi_{\pm}(r,Y)&=\frac{d}{d\alpha}\phi_{\pm}(r,Y,\alpha)\bigg|_{\alpha=0}\nonumber \\ 
&=-\sum_{n=1}^{\infty}(\mp \lambda)^n\int_{e^Y} ^{\infty}dx_1\int_0^{\infty}\prod_{i=2}^ndx_i \prod_{j=1}^n\frac{e^{-\frac{r}{2}\left(x_j+x_j^{-1}\right)}}{x_j+x_{j+1}} \ ,
\end{align}
we have
\begin{align}
{\cal Q}(r,Y)=\Phi_{+}(r,Y)-\Phi_{-}(r,Y) \ , \\
{\cal H}(r,Y)=\Phi_{+}(r,Y)+\Phi_{-}(r,Y) \ . 
\end{align}
Further more, we can write
\begin{align}\label{eq:decomY}
\Phi_{\pm}(r,Y)=\frac{\lambda}{2}\partial_{\lambda} \phi_{\pm}(r)+\Psi_{\pm}(r,Y) \ , 
\end{align}
where $\Psi_{\pm}(r,Y)$ are {\it odd} functions in $Y$
\begin{align} \label{eq:expformoddY}
\Psi_{\pm}(r,Y)&=\sum_{n=1}^{\infty}(\mp \lambda)^n\int_{1}^{e^{Y}} dx_1 \int_0^{\infty}\prod_{i=2}^ndx_i \prod_{j=1}^n\frac{e^{-\frac{r}{2}\left(x_j+x_j^{-1}\right)}}{x_j+x_{j+1}} \nonumber \\ 
&=\frac{1}{2}\sum_{n=1}^{\infty}(\mp \lambda)^n\int_{e^{-Y}}^{e^{Y}} dx_1 \int_0^{\infty}\prod_{i=2}^ndx_i \prod_{j=1}^n\frac{e^{-\frac{r}{2}\left(x_j+x_j^{-1}\right)}}{x_j+x_{j+1}} \ .
\end{align}
To show the above, simply change variables from $x_i \rightarrow \frac{1}{x_i}$. By taking the $\lambda$ derivatives, the $\lambda \partial_\lambda \varphi(r)$ and $\lambda \partial_\lambda \chi(r)$ themselves satisfy the linear differential equations Eq.~(\ref{eq:lineardifQ}) and Eq.~(\ref{eq:lineardifH}). Thus, the following functions which are odd in $Y$
\begin{align}
&\Psi(r,Y)=\Psi_+(r,Y)-\Psi_-(r,Y) \ , \label{eq:defPsi} \\
&\Xi(r,Y)=\Psi_+(r,Y)+\Psi_-(r,Y) \ ,  \label{eq:defXi}
\end{align}
also satisfy the Eq.~(\ref{eq:lineardifQ}) and Eq.~(\ref{eq:lineardifH}) with $\Psi$ replacing ${\cal Q}$, and $\Xi$ replacing ${\cal H}$. 

In the next section, we study the expansion of these functions in the small $r$ limit. More precisely, there are two limits which we consider: first, we consider $r \rightarrow 0$ with $Y$ fixed; second, we consider the scaling limit where $r \rightarrow 0$, $Y \rightarrow \infty$, while the scaling variable $t=\frac{r}{2}e^Y$ is fixed. As the case of the $Y$ independent functions, the physical value point $\pi\lambda=1$ is tricky in these limits. This is because that there are infinitely many contributions at $r^{j\left(1-\frac{2}{\pi}\arcsin \pi\lambda\right)}, \  j\in \mathbb{Z}_{>0}$ that are power corrections when $ 0<\lambda \pi<1$, but will collapse to logarithms when $\lambda\pi \rightarrow 1$. And the collapsed power corrections will be attached to all the contributions that remain to be ${\cal O}(r^2), {\cal O}(r^4).....$  when $\lambda \pi \rightarrow 1$. They must be resummed to obtain the correct expansion at $\pi \lambda=1$. On the other hand, the individual terms in the collapsed power corrections can be singular, only the full combination at any given power remain finite after cancellation of singularities. And similar to the $\epsilon\times\frac{1}{\epsilon}$ effect in perturbative calculations, logarithms that couple the short distance and large distance scales are generated at the end. 

The natural combination of the logarithm turns out to be~\cite{Wu:1975mw}
\begin{align} \label{eq:defomega}
\Omega=\ln \frac{8}{e^{\gamma_E}r} \ .
\end{align}
Here we stress that this is {\it not} the natural combination $-\frac{1}{2}\ln \frac{z^2m^2e^{2\gamma_E}}{4}$ that arises in the expansion of the free massive propagator $K_0(mz)$ and its derivatives. In particular, it is not the natural combination that could arise from any individual particle number's contribution in the form factor expansion. The difference in the transcendentality indicates that the spin field in the short distance limit remains far away from being Wick-polynomials of the free boson or fermion fields. In particular, one must sum over infinitely many particle numbers in order to obtain the correct short distance limit of the $\sigma$ correlators. The possibility of resuming the form factor expansion is mainly due to the following reasons. First, for $\pi \lambda<1$, the most singular part in the exponential form factor expansion is controlled by iterations of an integral operator and can be computed explicitly using related Mellin-Barnes representations.  Second, given the Mellin-Barnes computation as the boundary condition, the non-trivial Painleve III or Sinh-Gordon equation determines the power corrections and allows the resummation of the collapsed power corrections when $\lambda \pi \rightarrow 1$. In the next section we will see how such procedures can be generalized in the $(r,Y)$ two variable problem.  
\subsection{Conventions of the expansions}
In the rest of the work we will consider two types of short distance expansions of the various functions. To avoid confusions, it is helpful to establish the conventions of the expansions here.  It is useful to define the following function~\cite{McCoy:1976cd}
\begin{align}
\sigma\equiv\sigma(\lambda)=\frac{2}{\pi}\arcsin \pi\lambda \ ,  \  \lambda=\frac{1}{\pi}\sin \frac{\pi\sigma}{2} \ , \label{eq:defsigma}
\end{align}
and re-express all the $\lambda$ dependencies in terms of $\sigma$. The physical value limit $\lambda\pi \rightarrow 1$ then corresponds to $\sigma \rightarrow 1$ . The conventions of the expansions are then given as:
\begin{enumerate}
    \item The $r\rightarrow 0$, $Y={\cal O}(1)$ small-$r$ limit. In this limit, quantities are expressed using the variable list $(r,Y)$ and expanded as $r\rightarrow 0$ using the following convention (using $\Psi(r,Y)$ as an example)
    \begin{align}
    \Psi(r,Y)=\sum_{n=0}^{\infty}\Psi^{(n)}(r,Y) \ , 
    \end{align}
    where for sufficiently small $0<\epsilon<0.01$, at any $Y$ one has $r^{n+\epsilon}\le |\Psi^{(n)}(r,Y)| \le r^{n-\epsilon}$ as $\sigma \rightarrow 1$, $r\rightarrow  0$. The $\Psi^{(n)}(r,Y)$ includes all the power corrections that collapse to the power at $r^n$ when $\sigma\rightarrow 1$ and takes the form  $r^n a_n(r^{1-\sigma},Y,\sigma)$. After taking the $\sigma\rightarrow 1$ limit, the $a_n(r^{1-\sigma},Y,\sigma)$ becomes a rational function in the logarithm $\Omega$ defined in Eq.~(\ref{eq:defomega}), with $Y$-dependent coefficients.
    \item The $r\rightarrow 0$, $t=\frac{r}{2}e^Y={\cal O}(1)$ scaling limit. In this limit, quantities are changed to the variable list $(t,r)$ and expanded as $r \rightarrow 0$ using 
\begin{align}
    \Psi(t,r)=\sum_{n=0}^{\infty}\Psi^{(n)}(t,r) \ , 
    \end{align}
where $\Psi^{(n)}(t,r)$ denotes the contributions that collapse to the $r^n$ power as $\sigma \rightarrow 1$. It takes the form $\Psi^{(n)}(t,r)=r^nb_n(t,r^{1-\sigma},\sigma)$, where $b_n(t,r^{1-\sigma},\sigma)$ becomes a rational function in $\Omega$ with $t$-dependent coefficients as $\sigma \rightarrow 1$. 
\item For functions such as $\phi_{\pm}(r)$, $\eta(r)$ that are $Y$-independent, the two limits are identical and share the same convention. For the scaling functions $f_{\pm}(t)$ defined in Sec.~\ref{sec:scaling}, it is also useful to expand them at small $t$,  or equivalently, at small $r$ with fixed $Y$. We will use $f_{\pm}^{(n)}(t)$ to denote the contributions to $f_{\pm}^{(n)}(t)$ that collapse to the power $t^n$ as $\sigma\rightarrow 1$. Again, one can always write $f^{(n)}_{\pm}(t)=t^nc_{\pm}^{(n)}(t^{1-\sigma},\sigma)$ in which $c_{\pm}^{(n)}(t^{1-\sigma},\sigma)$ becomes a polynomial in $\ln t$ as $\sigma\rightarrow 1$. 
    \item The different expansions are distinguished according to the variable lists of the functions. When we write $f^{(n)}(r,Y)$ for an arbitrary function $f$ using the variable list $(r,Y)$, we always mean the contributions to $f$ that collapse to the power $r^n$, in the fixed $Y$, small-$r$ expansion. Similarly, $f^{(n)}(t,r)$ should always be interpreted under the fixed $t$, small-$r$ expansion. Notice that a function defined initially with a given variable list among $(r,Y)$ or $(t,r)$ can always be changed to another one and expanded in the limit correspondingly.

\end{enumerate}

\subsection{Summary of main results}\label{sec:summarymain}
For the reader's convenience, here we list the main results of the work.
\begin{enumerate}
    \item There are three scaling functions crucial to our analysis, $f_{\pm}(t)$ and $g(t)$. The first two $f_{\pm}(t)$ are given by the Barnes representations
\begin{align}
&f_{\pm}(t)=\int \frac{dz}{2\pi i}\int\frac{du}{2\pi i} t^{-z}{\cal M}_{\pm}(z,u) \ ,  \\ 
&{\cal M}_{\pm}(z,u)=\frac{(2 u-1) 2^{z-2} \sin \left(\frac{\pi  \sigma}{2}\right) \cos (\pi  u) \Gamma \left(\frac{z}{2}\right) \Gamma \left(\frac{2u+z-1}{2}\right) \Gamma \left(\frac{-2 u+z+1}{2} \right)}{\sqrt{\pi } z \Gamma \left(\frac{z+1}{2}\right) \left(\sin \left(\frac{\pi  \sigma}{2}\right)\pm\sin (\pi  u)\right)} \ ,
\end{align}
where the contour is chosen as $\frac{1}{2}<{\rm Re}(u)<1-\frac{\sigma}{2}$ and ${\rm Re}(z)>1-\sigma>2{\rm Re}(u)-1>0$. 
The last one $g(t)$ relates to $f_{\pm}(t)$ through
\begin{align}
g(t)=\frac{t^{\sigma-1}}{2b}\frac{2^{4 \sigma-5}  \sin \left(\frac{\pi  \sigma}{2}\right) \Gamma \left(\frac{\sigma-1}{2}\right)^2}{\pi ^2(1-\sigma)}+t^{\sigma-1}\int_0^{t}v^{-\sigma}(f_+(v)-f_-(v))dv \ ,
\end{align}
where $b$ is defined in Eq.~(\ref{eq:defB}). The $f_{\pm}(t)$ are introduced in Sec.~\ref{sec:scaling}, while $g(t)$ are introduced in Sec.~\ref{sec:scalinglimit}.
\item In the $\sigma \rightarrow 1$, these scaling functions are all finite and allow simple representations through Bessel $K$ functions. More precisely,  one has 
\begin{align}
&\lim_{\sigma \rightarrow 1^-} f_+(t)\equiv \mathfrak{f}_+(t)=\frac{t}{2\pi^2}K_0\left(\frac{t}{2}\right)K_1\left(\frac{t}{2}\right)-\frac{t^2}{4\pi^2}\left(K_1^2\left(\frac{t}{2}\right)-K_0^2\left(\frac{t}{2}\right)\right) \ ,  \\
&\lim_{\sigma \rightarrow 1^-} f_-(t)\equiv \mathfrak{f}_-(t)=\mathfrak{f}_+(t)-\frac{1}{\pi^2}K_0^2\left(\frac{t}{2}\right) \ , \\ 
& \lim_{\sigma \rightarrow 1^-}g(t) \equiv \mathfrak{g}(t)=-\frac{1}{\pi^2}\int_t^{\infty}\frac{K_0^2\left(\frac{v}{2}\right)}{v}dv \ . 
\end{align}
These simplifications were made manifest by investigating the CFT correlator on a finite cylinder in the appendix.~\ref{sec:cylinder}.  In Sec.~\ref{sec:scaling2}, we show how the $\mathfrak{f}_+(t)$ can be expressed as an integrated CFT correlator on the infinite plane, and in appendix.~\ref{sec:cylinder} we show how to generalize this to $\mathfrak{f}_-$ by going to a finite cylinder. 
\item Given the $f_{\pm}(t)$, their small-$t$ expansions that can be read from the Barnes representations. We can combine these information with the differential equations to determine the expansion of the number observables at the physical point $\sigma=1$. In the Sec.~\ref{sec:smallr}, we derive the small-$r$ expansions of the $N_{\pm}(r,Y)$ up to the third order at a fixed $r$. The results up to linear order in $r$ read
\begin{align}
&N_{\pm}(r,Y;\sigma=1)=\frac{1}{\pi^2}\left(\Omega-1-Y\right) \pm r\bigg(\frac{-2\Omega^3+\zeta_3-Y^3+3Y\Omega^2}{6\pi^2}\bigg)+{\cal O}(r^2)  \  .
\end{align}
Explicit formulas up to the third order can be found at Eq.~(\ref{eq:expandNsmallr}) and equations below.
\item In Sec.~\ref{sec:scalinglimit}, we derive the expansion of the number observable in the fixed $t$, small $r$ limit. The asymptotics formula is compact and involves the three scaling functions at $\sigma=1$: 
\begin{align}
N_{\pm}(t,r;\sigma=1)=\mathfrak{f}_{+}(t)\mp\frac{r}{2\pm r\Omega} \bigg( \left(\mathfrak{f}_+(t)-\mathfrak{f}_-(t)\right)\Omega+\mathfrak{g}(t)\bigg)+{\cal O}(r^4)  \ .
\end{align}
This is the major result of this work.  It is further tested against numerics in the appendix.~\ref{sec:numer}.

\end{enumerate}

\section{Scaling functions and asymptotics of the observables} \label{sec:analysis}
In this section, we begin to study the small-$r$ and large-$Y$ limits of the potential functions. For this purpose, it is useful to first review the small-$r$ asymptotics of the $\varphi(r)$ and $\chi(r)$ to see how singularities cancel when $\lambda\pi \rightarrow 1$. Let's first define the following functions~\cite{McCoy:1976cd}
\begin{align}
B\equiv B(\sigma)=2^{-3\sigma}\frac{\Gamma\left(\frac{1-\sigma}{2}\right)}{\Gamma \left(\frac{1+\sigma}{2}\right)} \ ,  b=-\frac{1}{16B^2(1-\sigma)^2} \  .\label{eq:defB} 
\end{align}
where $\sigma=\sigma(\lambda)$ is defined in Eq.~(\ref{eq:defsigma}). As $\sigma \rightarrow 1$, the $B(\sigma)$ develops an $\frac{1}{4(1-\sigma)}$ pole. In terms of the above, the leading power part in the $\sigma\rightarrow 1$ limit for $\varphi(r)$ and $\eta(r)$ can be given in closed forms~\cite{McCoy:1976cd}
\begin{align}
&\eta^{(1)}(r)=Br^{\sigma}\left(1+br^{2-2\sigma}\right) \ ,  \label{eq:eta0}\\
&\varphi^{(0)}(r)=-\ln B-\sigma\ln r-\ln \left(1+br^{2-2\sigma}\right) \ . \label{eq:phi0}
\end{align}
Corrections to the above are at least $r^4$ higher when $\sigma \rightarrow 1$. To obtain the above, one first notice that at small $r$, the individual contributions in the exponential form factor expansion Eq.~(\ref{eq:expformphipm}) are dominated by a constant and a single logarithm, which can be computed explicitly using the  Mellin-Barnes representations related to an integral operator. Then, after summing over $n$, one obtains the constants $B$ and $\sigma$. An important fact is that the sum over $n$ is absolutely convergent for $|\sigma|<1$, and the $o(1)$ corrections at the individual $n$ remain $o(1)$ after the summation~\cite{McCoy:1976cd,Tracy:1990tn}. Then, the Sinh-Gordon equation Eq.~(\ref{eq:SG1}) uniquely determines the structure of the power corrections at $|\sigma|<1$, and as $\sigma \rightarrow 1$, allows the resummation of the collapsed power corrections into closed forms. In particular, the power corrections to $\eta$ and $\varphi$ that collapse to the leading power are exactly given by Eq.~(\ref{eq:eta0}) and Eq.~(\ref{eq:phi0}).

To proceed further, it is interesting to notice that starting from the power $r^{2+2\sigma}$ and higher, the functions in $r^{2-2\sigma}$ resumming the collapsed power corrections can be obtained by solving linear differential equations.  In the Appendix.~\ref{sec:expansiondetail}, we illustrate our resummation procedure by deriving the next power corrections to $\eta$ and $\varphi$. As $\sigma \rightarrow 1$, these corrections maintain an ${\cal O}(r^4)$ gap to the leading order expressions Eq.~(\ref{eq:eta0}) and Eq.~(\ref{eq:phi0}). In deriving such power corrections, we will use the differential equations of the type Eq.~(\ref{eq:diffgen}) shown in the Appendix.~\ref{sec:diffalpha}. The results of the quartic corrections to $\eta$ and $\varphi$ read 
\begin{align}
&\eta^{(5)}(r)=Br^{5}\frac{B^2r^{3\sigma-3}}{16(1+\sigma)^2}\bigg(1-(1+2\sigma)w+\frac{(\sigma+1)^2(5-2\sigma)}{(\sigma-3)^2}w^2-\frac{(\sigma+1)^2}{(\sigma-3)^2}w^3\bigg) \ , \label{eq:etafourth}\\ 
&\varphi^{(4)}(r)=-\frac{\eta^{(5)}(r)}{Br^{\sigma}(1+b r^{2-2\sigma})} \ ,
\end{align}
where $w=-br^{2-2\sigma}$. It is then straightforward to show that in the $\sigma \rightarrow 1$ limit, the $\eta^{(1)}$, $\eta^{(5)}$, $\varphi^{(0)}$ and $\varphi^{(4)}$ are separately finite
\begin{align}
&\eta^{(1)}(r) \rightarrow \frac{r}{2}\Omega \ , \\
&\eta^{(5)}(r) \rightarrow \frac{r^5}{4096}(2 \Omega+1) \left(4 \Omega^2+2 \Omega+1\right) \ , 
\end{align}
where $\Omega$ is defined in Eq.~(\ref{eq:defomega}). The $\sigma\rightarrow 1$ limits above agree with the results in the literature~\cite{Wu:1975mw,McCoy:1976cd,Fonseca:2003ee}.

Given the $\varphi$, at any given power we must solve the linear equation Eq.~(\ref{eq:SG2}) with source term to determine $\chi$. The two solutions to the homogeneous equation are constant and $\ln r$, thus only affect the leading power. Although the coefficient of $\ln r$ can be fixed by the auxiliary equations Eq.~(\ref{eq:auxi1}), Eq.~(\ref{eq:auxi2}), the constant can not be fixed by the differential equations. The determination of this constant is called the ``Ising connecting problem'' in the literature and is finally solved in 1991~\cite{Tracy:1990tn}. The idea is still to sum the leading small-$r$ asymptotics at the individual-$n$ in the exponential form factor expansion Eq.~(\ref{eq:expformphipm}). The result of the $\chi(r)$ at the leading power reads~\cite{Tracy:1990tn}
\begin{align}
&\chi^{(0)}(r)+\varphi^{(0)}(r)=-\sigma\left(1-\frac{\sigma}{2}\right)\ln r+2\ln \tau_0 \ ,  \\ 
&\tau_0=\frac{\exp \left(\frac{3\sigma(2-\sigma)}{4}\ln 2\right)\Gamma_2\left(\frac{3-\sigma}{2}\right)\Gamma_2\left(\frac{1+\sigma}{2}\right)}{\Gamma_2 \left(\frac{3}{2}\right)\Gamma_2 \left(\frac{1}{2}\right)}  \ ,
\end{align}
where $\Gamma_2$ is the Barnes double gamma function.
Notice that $\tau_0=1$ when $\lambda=\sigma=0$. At the next two powers, $\chi $ can be completely determined from the differential equations as
\begin{align}
&\chi^{(2)}(r)=\frac{r^2}{8} \ , \\
&2\phi_+^{(4)}(r)=\chi^{(4)}(r)+\varphi^{(4)}(r)=-\frac{B^2}{16} r^{2+2\sigma} \left(\frac{2}{(\sigma+1)^2}+br^{2-2\sigma} +\frac{2 b^2 r^{4-4\sigma}}{(\sigma-3)^2}\right) \ . \label{eq:phiplusfourth}
\end{align}
In particular, in the $\sigma \rightarrow 1$ limit, one has
\begin{align}
\chi^{(4)}(r)+\varphi^{(4)}(r) \rightarrow -\frac{(8 \Omega (\Omega+1)+3) r^4}{1024} \ , 
\end{align}
which again agrees with the results in~\cite{Fonseca:2003ee}. Notice that the $\frac{r^2}{8}$ contribution at the quadratic power of $\chi^{(2)}$ is due to the constant term in Eq.~(\ref{eq:SG2}) and is $\sigma$ independent. 

Given the $\sigma$ dependent functions, one can determine the quantities $\lambda \partial_{\lambda} \phi_{\pm}$ that appear at $Y=0$ in Eq.~(\ref{eq:decomY}). They enter the observables $N_{\pm}$ through the equation Eq.~(\ref{eq:observable}). Up to the $r^3$, one only needs the $\lambda$ derivatives at the leading power. The result for the $\phi_+$ reads
\begin{align} \label{eq:derilambphilus}
\lim_{\sigma \rightarrow 1} \frac{\lambda\partial_{\lambda}\left(\chi(r)+\varphi(r)\right)}{2}=\frac{2}{\pi^2}(\Omega-1) +{\cal O}(r^4)\ . 
\end{align}
Notice the presence of the logarithm. On the other hand, one has
\begin{align}
\lim_{\sigma \rightarrow 1} \lambda \partial_{\lambda} \eta(r)=-\frac{r(2\Omega^3-\zeta_3)}{3\pi^2}+{\cal O}(r^5) \ . 
\end{align}
From the above, the $Y$ independent part of the $N_{\pm}(r,Y)$ can be separated out as
\begin{align}
N_{\pm}(r,Y)=\frac{1}{2}N_{\pm}(r)+\frac{{\Psi}(r,Y)+{\Xi }(r,Y)}{2}\mp\frac{\eta(r)}{1\pm\eta(r)}{\Psi}(r,Y) \ , 
\end{align}
where the odd functions $\Psi$ and $\Xi$ are defined in Eq.~(\ref{eq:defPsi}) and Eq.~(\ref{eq:defXi}), and the functions
\begin{align}
N_{\pm}(r)=\frac{\lambda\partial_{\lambda}\left(\chi(r)+\phi(r)\right)}{2}\pm\frac{1}{1\pm\eta(r)}\lambda\partial_{\lambda} \eta(r) \ ,
\end{align}
are exactly the averaged particle numbers underlying the two point function. At $\lambda \pi=1$, one has up to ${\cal O}(r^4)$ corrections, 
\begin{align}
N_{\pm}(r)=\frac{2 (\Omega-1)}{\pi ^2}\pm\frac{2}{2\pm r\Omega}\frac{r \left(-2 \Omega^3+\zeta_3\right)}{3 \pi ^2} +{\cal O}(r^4)\ . \label{eq:expandNpm}
\end{align}
Notice the double logarithmic enhancements in the power corrections. Furthermore, one can use the equations Eq.~(\ref{eq:etafourth}) and Eq.~(\ref{eq:phiplusfourth}) to further compute the corrections to the particle number beyond the ${\cal O}(r^4)$. Here we quote simply the $r^4$ contribution in Eq.~(\ref{eq:derilambphilus})
\begin{align}
\lim_{\sigma \rightarrow 1} \lambda\partial_\lambda \phi_+(r)=-\frac{r^4 \left(-32 \Omega^4-64 \Omega^3-72 \Omega^2+16 \Omega (\zeta_3-3)+(8 \zeta_3-15)\right)}{3072 \pi ^2} \ . 
\end{align}
Notice that the only transcendental number required here is the $\zeta_3$.  

\subsection{The scaling functions $f_{\pm}(t)$} \label{sec:scaling}
After introducing in details the rotational symmetric computations, let's consider the $Y$-dependent quantities. As a reminder, the crucial functions ${\cal Q}(r,Y)$ and ${\cal H}(r,Y)$ are controlled by the differential equations Eq.~(\ref{eq:lineardifQ}) and Eq.~(\ref{eq:lineardifH}). They allow exponential form factor expansions through the quantities $\Phi_{\pm}(r,Y)$ in Eq.~(\ref{eq:expformY}). Subtracting the $\frac{\lambda}{2}\partial_{\lambda}\phi_{\pm}(r)$, they can also be expressed in terms of the odd functions $\Psi_{\pm}(r,Y)$ given in Eq.~(\ref{eq:expformoddY}). We would like to study the small-$r$ limits of these quantities. 

The first observation that one can make, is that in the form factor expansions of $\Psi_{\pm}(r,Y)$, one can take the $r\rightarrow 0$ limit directly without $\ln r$ divergences at the leading power. This is due to the fact that the lower rapidity cutoff at $x_1=1$ removes the divergences when $x_i \rightarrow 0$, while the upper rapidity cutoff at $x_1=e^{Y}$ removes the divergences when $x_i \rightarrow \infty$. As such, as $r \rightarrow 0$, for $0<\sigma<1$ one has the following
\begin{align}
\Psi_{\pm}(r,Y)=\sum_{n=1}^{\infty}(\mp \lambda)^n\int_{1}^{e^{Y}} dx_1 \int_0^{\infty}\prod_{i=2}^ndx_i \prod_{j=1}^n\frac{1}{x_j+x_{j+1}}+o(1) \ . 
\end{align}
It is not hard to show that one has
\begin{align}
\int_{1}^{e^{Y}} dx_1 \int_0^{\infty}\prod_{i=2}^ndx_i \prod_{j=1}^n\frac{1}{x_j+x_{j+1}}=Y\int_0^{\infty}\frac{dx_1...dx_{n-1}}{(1+x_1)(x_1+x_2)...(x_{n-1}+1)} \ .
\end{align}
While the last integral is exactly the one for the $\ln r$ term of $\phi_{\pm}$:
\begin{align}
&\int_0^{\infty}\frac{dx_1...dx_{n-1}}{(1+x_1)(x_1+x_2)...(x_{n-1}+1)}=\int_0^{\infty} d\rho e^{-\rho(x_1+1)} \frac{dx_1..dx_{n-1}}{(x_1+x_2)....(x_{n-1}+1)} \nonumber \\ 
&=\int_0^{\infty} d\rho e^{-\alpha_1-\rho} \frac{d\alpha_1..d\alpha_{n-1}}{(\alpha_1+\alpha_2)....(\alpha_{n-1}+\rho)}=\frac{\pi^{n-1}}{2}\int_{-\infty}^{\infty} dy \frac{1}{\cosh^n \pi y} \equiv J_n \ . 
\end{align}
Since in the small-$r$ expansion of the $\phi_{\pm}(r)$, one has~\cite{McCoy:1976cd,Tracy:1990tn}
\begin{align}
&\phi_{\pm}(r)=- \sigma_{\pm}(\lambda)\ln r+\beta_{\pm}(\lambda)+o(1) \  , \label{eq:phpms}  \\ 
&\sigma_{\pm}=-2\sum_{n=1}^{\infty}\lambda^n\frac{(\mp 1)^n}{n}J_n =\frac{\pm\sigma}{2}\left(1\mp\frac{\sigma}{2}\right)\ , 
\end{align}
one obtains the following leading small-$r$ asymptotics
\begin{align}
\Psi_{\pm}(r,Y)=-\frac{\lambda}{2}\partial_{\lambda} \sigma_{\pm}(\lambda) Y+o(1) \ . \label{eq:PsiasymL}
\end{align}
Notice that as $\sigma \rightarrow 1$, the $\lambda$ derivative remains finite for $\sigma_+$ , but diverges for $\sigma_-$. Clearly, as $\sigma \rightarrow 1$, there will be many power corrections in the $o(1)$ terms that will collapse, similar to what happened to $\phi_{\pm}$. One must find out a proper way to determine the $o(1)$ terms in Eq.~(\ref{eq:PsiasymL}) and resum the collapsed contributions. 

To search for the $o(1)$ corrections, one might hope to follow the approach in the $Y$ independent version, namely, to use the equations Eq.~(\ref{eq:lineardifQ}) and Eq.~(\ref{eq:lineardifH}) to fix these contributions. The first observation is that the $-\frac{\lambda}{2}\partial_{\lambda} \sigma_{\pm}(\lambda) Y$ term in Eq.~(\ref{eq:PsiasymL}), when propagates to the high-powers, can be separated from the rest of the solution:
\begin{align}
\Psi_{\pm}(r,Y)=\frac{\lambda}{2}\partial_{\lambda} \phi_{\pm}\bigg|_{\ln r} Y+V_\pm(r, Y) \ , \label{eq:decompPsi}
\end{align}
where $A|_{\ln r}$ denotes the coefficient of $\ln r$ in the expression $A$, which can be a function in $r$. The point is that the $\lambda \partial_\lambda \phi_{\pm}$ satisfies the second order linear equations Eq.~(\ref{eq:lineardifQ}) and Eq.~(\ref{eq:lineardifH}) in which there are no logarithms in the coefficients, thus the $\ln r$ is proportional to the solution without logarithms.

On the other hand, the functions $V_{\pm}(r, Y)$ contain only powers in $r$ and hyperbolic-sins in $Y$. The combinations
\begin{align}
V(r, Y)=V_+(r, Y)-V_-(r, Y) \ , \ Z(r, Y)=V_+(r, Y)+V_-(r, Y) \ , 
\end{align}
continue to satisfy the linear equations Eq.~(\ref{eq:lineardifQ}) and Eq.~(\ref{eq:lineardifH}), with $V$ replacing the ${\cal Q}$, while $Z$ replacing the ${\cal H}$. However, unlike the one-variable version, at any given power of $r$, there remains a second-order differential equation for the $Y$ dependencies. Since the $\Psi_{\pm}$ and $V_{\pm}$ are odd in $Y$, the unknown constants can be reduced to $1$, but can not be further reduced. And these constants can not be determined by the differential equations themselves. At the end, at each power $r^a$ where $a=a(\sigma)>0$, there will be a new constant $c_a$ to be fixed, which contributes to the $\Psi$ in the form  $c_ar^a \sinh aY$. The integration constants at lower powers will then propagate to higher powers as inhomogeneous terms, and contribute to $r^b\sinh aY$ terms with $b>a$. 

By investigating the individual form-factor contributions, one notice that there are linear power corrections for $Y \ne 0$ at the fixed particle numbers. As such, we expect the first $o(1)$ contribution in Eq.~(\ref{eq:PsiasymL}) to be at the power $r^{1-\sigma}$. The $Y$ dependency of $V(r,Y)$ at this power must be given by $\sinh (1-\sigma)Y$. This term, when multiplied by the leading-power part of $\cosh 2\phi\sim \eta^{-2}\sim B^{-2}r^{-2\sigma}(1+br^{2-2\sigma})^{-2}$ in the linear equation Eq.~(\ref{eq:lineardifQ}), will then propagate to higher powers at $r^{1-\sigma+j(2-2\sigma)}$. Denote the contributions to $V(r,Y)$ at the power $r^{j(1-\sigma)}$ by $r^{j(1-\sigma)}f_j(Y)$, then the equation for $f_j$ takes the following form
\begin{align}
j^2(1-\sigma)^2f_j(Y)-f^{''}_j(Y)=\sum_{k<j}c_k\sinh k(1-\sigma)Y \ ,
\end{align}
where the right-hand side is exactly due to the source terms propagated from the lower-powers. The solution odd in $Y$ still contains an unknown constant for the homogeneous equation. To determine these constants, it is sufficient to know the largest term in the {\it scaling limit} when $r \rightarrow 0,  \ Y \rightarrow \infty$ while $re^{Y}$ remains fixed. Indeed, in this limit, all contributions which are due to source terms from the lower powers are sub-leading, and one exactly probes the constants multiplying the solution to the homogeneous equations. 

Indeed, from Eq.~(\ref{eq:expformY}), it is clear that in the scaling limit, the largest contributions in the form factor integrals can be obtained by dropping the $x_j^{-1}$ contribution in the exponential form factor expansion:
\begin{align}
&\int_{e^Y} ^{\infty}dx_1\int_0^{\infty}\prod_{i=2}^ndx_i \prod_{j=1}^n\frac{e^{-\frac{r}{2}\left(x_j+x_j^{-1}\right)}}{x_j+x_{j+1}}=\int_{1} ^{\infty}dx_1\int_0^{\infty}\prod_{i=2}^ndx_i \prod_{j=1}^n\frac{e^{-\frac{r}{2}\left(e^{Y}x_j+e^{-Y}x_j^{-1}\right)}}{x_j+x_{j+1}} \nonumber \\ 
&\rightarrow \int_{1} ^{\infty}dx_1\int_0^{\infty}\prod_{i=2}^ndx_i \prod_{j=1}^n\frac{e^{-\frac{r}{2}e^Yx_j}}{x_j+x_{j+1}}=\int_{\frac{r}{2}e^Y} ^{\infty}dx_1\int_0^{\infty}\prod_{i=2}^ndx_i \prod_{j=1}^n\frac{e^{-x_j}}{x_j+x_{j+1}} \ .
\end{align}
The integral is absolutely convergent for $r>0$. After summing over the $n$, one obtains the following {\it scaling functions} 
\begin{align} \label{eq:defscaling}
f_{\pm}(t)=-\sum_{n=1}^{\infty}(\mp \lambda)^n\int_{t} ^{\infty}dx_1\int_0^{\infty}\prod_{i=2}^ndx_i \prod_{j=1}^n\frac{e^{-x_j}}{x_j+x_{j+1}} \ , \  t=\frac{r}{2}e^{Y} \ . 
\end{align}
Expand the scaling functions in the small $t$ limit, one obtains the desired largest contributions in $Y$ of the form $(re^{Y})^a$ at each power in the small $r$ expansion, when $0<\sigma<1$. In the rest of this subsection, we show how the sum over $n$ of for the scaling functions can be performed to obtain the expansion in the small-$t$ limit. We will first work with $0<\sigma<1$ situation, then consider the subtle $\sigma \rightarrow 1$ limit.

To proceed with the scaling functions in Eq.~(\ref{eq:defscaling}), we introduce the integral operator~\cite{McCoy:1976cd,Tracy:1990tn,Marino:2021dzn}
\begin{align}
(\hat Af)(x)=\int_0^{\infty} \frac{e^{-(x+y)/2}f(y)}{x+y}dy \ , 
\end{align}
which is bounded and self-adjoint on $L^2(\mathbb{R}_+)$. Its spectrum is purely continuous. The eigenvalues and normalized eigen-functions are labeled by a continuous variable $p\ge 0$, which can be given in closed forms~\cite{McCoy:1976cd,Tracy:1990tn,Marino:2021dzn}
\begin{align}
&\hat Af_p(x)=\lambda_p f_p(x) \ , \lambda_p=\frac{\pi}{\cosh \pi p} \in[0,\pi]\ ,  \\
&f_p(x)=\frac{\sqrt{2p\sinh \pi p}}{\pi}\frac{1}{\sqrt{x}}K_{ip}\left(\frac{x}{2}\right) \ , \\
&\int_0^{\infty} dx f_p(x)f_{p'}(x)=\delta(p-p') \ .
\end{align}
Given the integral operator, it is clear that the form factor integrals of the scaling functions can be expressed as iterations of the integral operator
\begin{align}
    I_n(t)&=\int_{t} ^{\infty}dx_1\int_0^{\infty}\prod_{i=2}^ndx_i \prod_{j=1}^n\frac{e^{-x_j}}{x_j+x_{j+1}}\nonumber \\ 
    &=\int_{t}^{\infty} dx_1 \langle x_1|\hat A^n|x_1\rangle=\frac{1}{2}\int_{-\infty}^{\infty} \lambda_p^ndp \int_t^{\infty} f_p(x_1)^2dx_1 \ . 
\end{align}
To proceed, we perform the Mellin transform with respect to $t$
\begin{align}
\int_0^{\infty} dt t^{z-1}\int_t^{\infty}f_p^2(x_1) dx_1=\frac{2p\sinh \pi p}{\pi^2z}\int_0^{\infty }x^{z-1}K^2_{ip}\left(\frac{x}{2}\right) dx \ .
\end{align}
The last integral can be integrated into gamma function using the standard two-$\Gamma$ Mellin-Barnes representation of the Bessel-$K$ function and the first Barnes lemma. This leads to
\begin{align}
I_n(z)=\int_0^{\infty} dt t^{z-1}I_n(t)=\int_{-\infty}^{\infty}\frac{dp}{2\pi}\lambda_p^n \frac{2^{z-1} p\sinh (\pi  p) \Gamma \left(\frac{z}{2}\right) \Gamma \left(\frac{z}{2}-i p\right) \Gamma \left(\frac{z}{2}+ip\right)}{\sqrt{\pi}z \Gamma \left(\frac{z+1}{2}\right)} \ .
\end{align}
Then, after summing over $n$ which is convergent for $-1<\sigma<1$, one obtains
\begin{align}
&{\cal M}(f_{\pm})(z)=-\sum_{n=1}^{\infty}(\mp \lambda)^nI_n(z)\nonumber \\ 
&=\int_{-\infty}^{\infty}\frac{dp}{2\pi}\frac{2^{z-1} p\sinh (\pi  p) \Gamma \left(\frac{z}{2}\right) \Gamma \left(\frac{z}{2}-i p\right) \Gamma \left(\frac{z}{2}+ip\right)}{\sqrt{\pi}z \Gamma \left(\frac{z+1}{2}\right)}\frac{\sin \frac{\pi \sigma}{2}}{\pm \cosh \pi p+\sin\frac{\pi \sigma}{2}} \ .
\end{align}
The above suggest to introduce the Barnes parameter $u=\frac{1}{2}+ip$, and write
\begin{align}
{\cal M}(f_{\pm})(z)=\int\frac{du}{2\pi i}{\cal M}_{\pm}(z,u) \ , 
\end{align}
where
\begin{align}
&{\cal M}_{\pm}(z,u)=\frac{(2 u-1) 2^{z-2} \sin \left(\frac{\pi  \sigma}{2}\right) \cos (\pi  u) \Gamma \left(\frac{z}{2}\right) \Gamma \left(\frac{2u+z-1}{2}\right) \Gamma \left(\frac{-2 u+z+1}{2} \right)}{\sqrt{\pi } z \Gamma \left(\frac{z+1}{2}\right) \left(\sin \left(\frac{\pi  \sigma}{2}\right)\pm\sin (\pi  u)\right)} \ . \label{eq:BarneGrandscaling}
\end{align}
In terms of the above, one obtains the important Mellin-Barnes representation for the $f_{\pm}(t)$
\begin{align}
f_{\pm}(t)=\int \frac{dz}{2\pi i}\int\frac{du}{2\pi i} t^{-z}{\cal M}_{\pm}(z,u) \ . \label{eq: MBscaling}
\end{align}
Here the contour of the Barnes integral can be chosen as $\frac{1}{2}<{\rm Re}(u)<1-\frac{\sigma}{2}$ and ${\rm Re}(z)>1-\sigma>2{\rm Re}(u)-1>0$. In particular, for the $f_-$, the contour is pinched between the $u_1=\frac{\sigma}{2}$ and $u_2=1-\frac{\sigma}{2}$ poles. However, we will see that both of the $f_{\pm}(t)$ will remain finite as $\sigma \rightarrow 1$. Moreover, we will show later that $f_+(t)$ at $\sigma=1$ can be written as an integral of the $\langle\sigma \psi \psi \sigma\rangle/\langle\sigma\sigma\rangle$ four-point function ($\sigma$ means the spin-operator here) in the Ising CFT, and $f_-(t)$ can be expressed as integrals of $f_+(t)$. 

Given the crucial Mellin-Barnes representations, one can obtain the small-$t$ expansion of the scaling functions by shifting the $z$ to the left and pick up the poles. It is easy to see that at the leading power, one needs the following poles from ${\cal M}_+$: $z=0$; $z=2u-1$, then $u=\frac{1}{2}$. This leads to the leading power part of the $f_+$
\begin{align}
f_+^{(0)}(t)=&\int \frac{du}{2\pi i}\frac{\sin \left(\frac{\pi  \sigma}{2}\right) \left(-2 \ln t+4\ln 2+\psi \left(\frac{1}{2}-u\right)+\psi\left(u-\frac{1}{2}\right)\right)}{2 \left(\sin \left(\frac{\pi  \sigma}{2}\right)+\sin (\pi  u)\right)} \nonumber \\ 
&-\frac{\sin \left(\frac{\pi  \sigma}{2}\right)}{4 \left(\sin \left(\frac{\pi  \sigma}{2}\right)+1\right)} \ .
\end{align}
Similarly, the leading power part of $f_-$ can be determined from the following poles: $z=0$; $z=2u-1$, then $u=\frac{1}{2}$ or $u=\frac{\sigma}{2}$; $z=-2u+1$, then (to the right) $u=1-\frac{\sigma}{2}$. The result reads
\begin{align}
f_-^{(0)}(t)=&\int \frac{du}{2\pi i}\frac{\sin \left(\frac{\pi  \sigma}{2}\right) \left(-2 \ln t+4\ln 2+\psi \left(\frac{1}{2}-u\right)+\psi\left(u-\frac{1}{2}\right)\right)}{2 \left(\sin \left(\frac{\pi  \sigma}{2}\right)-\sin (\pi  u)\right)} \nonumber \\ 
&-\frac{\sin \left(\frac{\pi  \sigma}{2}\right)}{4 \left(\sin \left(\frac{\pi  \sigma}{2}\right)-1\right)}-\frac{2^{2 \sigma-3} t^{1-\sigma} \sin \left(\frac{\pi  \sigma}{2}\right) \Gamma \left(\frac{\sigma-1}{2}\right)^2}{\pi ^2} \ .
\end{align}
Furthermore, power corrections to them are ${\cal O}(t^2)$ as $\sigma \rightarrow 1$. As expected, contributions at the power $r^{1-\sigma}$ appear. Here we show how the remaining Barnes integrals above can be related to those appeared in the Ising connecting problem.  First, for the logarithmic term, one has
\begin{align}
\int \frac{du}{2\pi i} \frac{\sin \frac{\pi \sigma}{2}}{\sin \frac{\pi \sigma}{2}\pm\sin \pi u}=\pm\frac{1\mp\sigma}{2\pi}\frac{\sin \frac{\pi \sigma}{2}}{\cos \frac{\pi\sigma}{2}} \equiv \frac{1}{2}\lambda\partial_\lambda \sigma_{\pm} \ . 
\end{align}
This is consistent with the expectation to the coefficients of the log-term. So we need to consider the constant terms
\begin{align}
I_{\pm}(\sigma)=\sin \frac{\pi\sigma}{2}\int \frac{du}{4\pi i}\frac{\psi \left(\frac{1}{2}-u\right)+\psi\left(u-\frac{1}{2}\right)+6\ln 2}{ \sin \left(\frac{\pi  \sigma}{2}\right) \pm \sin (\pi  u)} -\frac{\sin \frac{\pi \sigma}{2}}{4 \left(\sin \frac{\pi\sigma}{2}\pm 1\right)}\ . 
\end{align}
Now, by expanding the integrals again, one has
\begin{align}
I_{\pm}(\sigma)=-\sum_{n=1}^{\infty}(\mp\lambda)^n\bigg(3\ln 2 J_n+\int_{-\infty}^{\infty}\frac{dp}{2\pi}\left(\frac{\pi}{\cosh \pi p}\right)^n\psi(i p)-\frac{\pi^n}{4}\bigg)=\frac{1}{2}\lambda \partial_{\lambda} \beta_{\pm} \ . 
\end{align}
Here, we have used the fact that $\beta_{\pm}(\lambda)$ are given by summing the $\beta_k$ in Eq.~(3.10) of~\cite{Tracy:1990tn}. As such, we have
\begin{align}
&f_{+}^{(0)}(t)=\frac{1}{2}\lambda\partial_\lambda \bigg(\sigma_{+} \ln \frac{1}{2t}+\beta_+(\lambda) \bigg) \ , \label{eq:f0plus}\\
&f_{-}^{(0)}(t)=\frac{1}{2}\lambda\partial_\lambda \bigg(\sigma_{-} \ln \frac{1}{2t}+\beta_-(\lambda) \bigg)-\frac{2^{2 \sigma-3} t^{1-\sigma} \sin \left(\frac{\pi  \sigma}{2}\right) \Gamma \left(\frac{\sigma-1}{2}\right)^2}{\pi ^2} \ . \label{eq:f0minus}
\end{align}
It its interesting to notice that all the contributions at ${\cal O}(r^{0})$ are included in the scaling function, including the leading asymptotics of the $\phi_{\pm}$ in the Eq.~(\ref{eq:phpms}). On the other hand, the $f_{-}^{(0)}(t)$ also contains a $t^{1-\sigma}$ term that will collapse to the leading power as $\sigma \rightarrow1$. It is not hard to show that it exactly cancels the singularities in the first term, and the full results of the $f^{(0)}_{\pm}(t)$ remain finite as $\sigma \rightarrow 1$:
\begin{align}
&\lim_{\sigma\rightarrow 1} f_+^{(0)}(t)= \frac{L-1}{\pi^2} \ , \label{eq:fplus0pi}\\
&\lim_{\sigma \rightarrow 1} f_{-}^{(0)}(t)=\frac{L-1}{\pi^2}-\frac{L^2}{\pi^2} \ , 
\end{align}
where we have introduced the following logarithms in the scaling variable
\begin{align}
L=\ln \frac{4}{e^{\gamma_E}t} \ . \label{eq:defL}
\end{align}
Latter we will use the Eq.~(\ref{eq:f0minus}) in order to determine the full-$Y$ dependencies of $V_{\pm}(r,Y)$ at the leading power.

By shifting the $z$-contours further to the left, the ${\cal O}(t^2)$ power corrections to the scaling functions can also be systematically determined. For the $f_+(t)$, we need the following poles: $z=-2$; $z=2u-1$ then $u=-\frac{1}{2}, u=-\frac{\sigma}{2},u=-1+\frac{\sigma}{2}$; $z=-2u+1$ then $u=\frac{3}{2}, u=1+\frac{\sigma}{2}, u=2-\frac{\sigma}{2}$. And for the $f_-(t)$, we need the following: $z=-2$; $z=2u-1$ then $u=-\frac{1}{2}$; $z=-2u+1$ then $u=\frac{3}{2}$; $z=2u-3$ then $u=\frac{\sigma}{2}$; $u=-2u-1$ then $u=1-\frac{\sigma}{2}$. By picking up these poles, we obtains the following Barnes representations
\begin{align}
&f_+^{(2)}(t)=\int\frac{du}{2\pi i}\frac{t^2 \sin \left(\frac{\pi  \sigma}{2}\right)}{4 \left(4 u^2-4 u-3\right) \left(\sin \left(\frac{\pi  \sigma}{2}\right)+\sin (\pi  u)\right)}+\frac{t^2 \sin \left(\frac{\pi  \sigma}{2}\right)}{32 \left(\sin \left(\frac{\pi  \sigma}{2}\right)-1\right)}\nonumber \\ 
&+\frac{2^{\sigma-3} t^{3-\sigma} \sin \left(\frac{\pi  \sigma}{2}\right) \Gamma (\sigma-3) \Gamma \left(\frac{\sigma-3}{2}\right)}{\pi ^{3/2} \Gamma \left(\frac{\sigma}{2}-1\right)}+\frac{2^{-\sigma-1} t^{\sigma+1} \sin \left(\frac{\pi  \sigma}{2}\right) \Gamma (-\sigma-1) \Gamma \left(-\frac{\sigma+1}{2} \right)}{\pi ^{3/2} \Gamma \left(-\frac{\sigma}{2}\right)} \ .
\end{align}
And 
\begin{align}
&f_-^{(2)}(t)=\int\frac{du}{2\pi i}\frac{t^2 \sin \left(\frac{\pi  \sigma}{2}\right)}{4 \left(4 u^2-4 u-3\right) \left(\sin \left(\frac{\pi  \sigma}{2}\right)-\sin (\pi  u)\right)}+\frac{t^2 \sin \left(\frac{\pi  \sigma}{2}\right)}{32 \left(\sin \left(\frac{\pi  \sigma}{2}\right)+1\right)}\nonumber \\ 
&+\frac{2^{\sigma-3} (\sigma-1) t^{3-\sigma} \sin \left(\frac{\pi  \sigma}{2}\right) \Gamma \left(\frac{\sigma-3}{2}\right) \Gamma (\sigma-2)}{\pi ^{3/2} (\sigma-3) \Gamma \left(\frac{\sigma}{2}-1\right)} \ .
\end{align}
The Barnes integrals here are much simpler than at the leading power, since the digamma functions are absent, and the $1/u^2$ suppression allows them to be evaluated by summing up the residues.  The results read 
\begin{align}
&f^{(2)}_+(t)=\frac{t^2 \tan \left(\frac{\pi  \sigma}{2}\right)}{8 \pi  (\sigma-1)}\nonumber \\ 
&+\frac{2^{\sigma-3} t^{3-\sigma} \sin \left(\frac{\pi  \sigma}{2}\right) \Gamma (\sigma-3) \Gamma \left(\frac{\sigma-3}{2}\right)}{\pi ^{3/2} \Gamma \left(\frac{\sigma}{2}-1\right)}+\frac{2^{-\sigma-1} t^{\sigma+1} \sin \left(\frac{\pi  \sigma}{2}\right) \Gamma (-\sigma-1) \Gamma \left(-\frac{\sigma+1}{2}\right)}{\pi ^{3/2} \Gamma \left(-\frac{\sigma}{2}\right)} \ , \label{eq:fplus2}
\end{align}
and
\begin{align}
f_-^{(2)}(t)=\frac{t^2 \tan \left(\frac{\pi  \sigma}{2}\right)}{8 \pi  (\sigma+1)}+\frac{2^{\sigma-3} (\sigma-1) t^{3-\sigma} \sin \left(\frac{\pi  \sigma}{2}\right) \Gamma \left(\frac{\sigma-3}{2}\right) \Gamma (\sigma-2)}{\pi ^{3/2} (\sigma-3) \Gamma \left(\frac{\sigma}{2}-1\right)} \ . \label{eq:fminus2}
\end{align}
As such, in all the contributions that will collapse to the quadratic power when $\sigma \rightarrow 1$, there are three powers: $r^2$, $r^{3-\sigma}$, $r^{\sigma+1}$ that will be attached to the same number in $e^Y$. This will be used later to determine the $r^2$ corrections in $V_{\pm}(r, Y)$. Notice that the $t^{1+\sigma}$ term in the $f_+^{(2)}(t)$ simply relates to the $t^{1-\sigma}$ term of $f_{-}^{(0)}(t)$ through $\sigma\rightarrow-\sigma$, and the $t^2$ terms in the $f_{\pm}^{(2)}(t)$ also relate in the same way, reflecting the $\lambda\leftrightarrow-\lambda$ relationship between the two scaling functions. On the other hand, since as $\sigma\rightarrow1$, flipping the sign of $\sigma$ will clearly change the status in the small-$t$ expansion, it is more helpful for our purpose to treat them as two scaling functions instead of one. We also notice that the $\sigma \rightarrow 1$ limits are again finite
\begin{align}
&\lim_{\sigma \rightarrow 1} f_+^{(2)}(t)=\frac{t^2}{16\pi^2}(2L^2+4L+3) \ , \\
&\lim_{\sigma \rightarrow 1} f_-^{(2)}(t)=\frac{t^2}{16\pi^2}(3+2L) \ . 
\end{align}
The finiteness in the $\sigma \rightarrow1$ limit of the scaling function will persists to all powers in $t$, as we will show next. 
\subsection{The scaling functions at $\sigma=1$} \label{sec:scaling2}
In this subsection we study the $\sigma\rightarrow 1$ limit of the scaling functions. The finiteness of the $f_{+}(t)$ at $\sigma=1$ is clear from Eq.~(\ref{eq:BarneGrandscaling}). Thus we have
\begin{align}
\lim_{\sigma\rightarrow1}f_+(t)=\int\frac{dz}{2\pi i}t^{-z}\int\frac{du}{2\pi i}\frac{2^{z-2}\Gamma \left(\frac{z}{2}\right)(2 u-1)  \cos (\pi  u)  \Gamma \left(\frac{2u+z-1}{2}\right) \Gamma \left(\frac{-2u+z+1}{2}\right)}{\sqrt{\pi } z \Gamma \left(\frac{z+1}{2}\right)(\sin (\pi  u)+1) } \ . 
\end{align}
For the $f_{-}(t)$, this is not manifest at the first glance, as the $\frac{\sigma}{2}$ and $1-\frac{\sigma}{2}$ poles are pinched. To separate the possible singularities, one can shift the $u$ to the right, between $\frac{1+1-\sigma}{2}<u<\frac{1+z}{2}$. Then we can take the $\sigma=1$ limit in the Barnes integral with the modified $u$ path, as the singularity no-longer pinch. All singularities are then encoded by the residue of ${\cal M}_-(z,u)$ at $u=1-\frac{\sigma}{2}$. However, we have
\begin{align}
-{\rm Res}_{u=1-\frac{\sigma}{2}} {\cal M}_-(z,u)=-\frac{(\sigma-1) 2^{z-2} \sin \left(\frac{\pi  \sigma}{2}\right) \Gamma \left(\frac{z}{2}\right) \Gamma \left(\frac{-\sigma+z+1}{2} \right) \Gamma \left(\frac{\sigma+z-1}{2} \right)}{\pi ^{3/2} z \Gamma \left(\frac{z+1}{2}\right)} \ .
\end{align}
This term is not only non-singular, but vanishes in the $\sigma \rightarrow 1$ limit. As such, one obtains the following Mellin representation of $f_{-}(t)$
\begin{align}
\lim_{\sigma\rightarrow1}f_-(t)=\int\frac{dz}{2\pi i}t^{-z}\int\frac{du}{2\pi i}\frac{2^{z-2}\Gamma \left(\frac{z}{2}\right)(2 u-1)  \cos (\pi  u)  \Gamma \left(\frac{2u+z-1}{2}\right) \Gamma \left(\frac{-2u+z+1}{2}\right)}{\sqrt{\pi } z \Gamma \left(\frac{z+1}{2}\right)(1-\sin (\pi  u)) } \ ,
\end{align}
where the contour is now $\frac{1}{2}<u<1$ and $z>2u-1$. The task now is to further simplify the $du$ Barnes integrals. We introduce 
\begin{align}
d_{\pm}(z)=&\int\frac{du}{2\pi i}\frac{(2 u-1)  \cos (\pi  u)  \Gamma \left(\frac{2u+z-1}{2}\right) \Gamma \left(\frac{-2u+z+1}{2}\right)}{1\pm\sin \pi u } \ , \nonumber \\ 
=-&\int\frac{ds}{2\pi i}\frac{2s\sin(\pi  s)  \Gamma \left(s+\frac{z}{2}\right) \Gamma \left(-s+\frac{z}{2}\right)}{1\pm\cos \pi s }  \ ,
\end{align}
where we have $\frac{1}{2}<{\rm Re}(u)< {\rm min} \{\frac{1+z}{2},\frac{3}{2}\}$, ${\rm Re}(z)>0$, and in the last line we changed the variable to $s=u-\frac{1}{2}$. Here we claim the following:
\begin{align}
d_+(z)=\frac{2 \Gamma \left(\frac{z}{2}+1\right)^2}{\pi  (z+1)} \ , \ d_-(z)=-\frac{z+2}{z}d_+(z) \ .  \label{eq:identity}
\end{align}
These identities can be established in the following manners. First, the addition $d_+(z)+d_-(z)$ can be simplified as
\begin{align}
&d_+(z)+d_-(z)=-\int\frac{ds}{2\pi i}\frac{4s \Gamma \left(s+\frac{z}{2}\right) \Gamma \left(-s+\frac{z}{2}\right)}{\sin \pi s } \nonumber \\ 
&=-\frac{4}{\pi}\int\frac{ds}{2\pi i}\Gamma(1+s)\Gamma(1-s) \Gamma \left(s+\frac{z}{2}\right) \Gamma \left(-s+\frac{z}{2}\right) \ . 
\end{align}
As such, it can be evaluated using the first Barnes lemma into products and ratios of gamma functions. Second, the difference can be evaluated as
\begin{align}
d_+(z)-d_-(z)=\lim_{x \rightarrow 1^-}\int\frac{ds}{2\pi i}\frac{4s \cos\pi s\Gamma \left(s+\frac{z}{2}\right) \Gamma \left(-s+\frac{z}{2}\right)}{\sin \pi s } x^{-s} \ ,
\end{align}
where we have introduced a suppression factor $x^{-s}$  with $0<x<1$. It can then be evaluated by shifting the contour to the left and picking up the residues. There are two series of poles. The $s=-k$ series leads to a $_{2}F_1$-type hypergeometric series, while the $s=-\frac{z}{2}-k$ leads to the small-$x$ expansion of the $(1+x)^{-z}$ type algebraic functions. They can be summed and after sending $x\rightarrow 1$, expressed in terms of gamma functions. Combining the $d_+(z) \pm d_-(z)$ one obtains the crucial simplifications Eq.~(\ref{eq:identity}).

Given Eq.~(\ref{eq:identity}), the scaling functions $\mathfrak{f}_{\pm}(t) \equiv\lim_{\sigma\rightarrow 1}f_{\pm}(t)$ can be represented as one variable Barnes integrals
\begin{align}
\mathfrak{f}_{+}(t)=\int \frac{dz}{2\pi i}t^{-z}\frac{2^{z} z \Gamma \left(\frac{z}{2}\right)^3}{16\pi ^{3/2} \Gamma \left(\frac{z+3}{2}\right)} \ , \  \mathfrak{f}_{-}(t)=-\int \frac{dz}{2\pi i}t^{-z}\frac{2^{z} (z+2) \Gamma \left(\frac{z}{2}\right)^3}{16\pi ^{3/2} \Gamma \left(\frac{z+3}{2}\right)} \ , \label{eq:scalingsimple}
\end{align}
where ${\rm Re}(z)>0$. From the above,  the following relation
\begin{align} \label{eq:fminusdif}
t\mathfrak{f}_+'(t)+t\mathfrak{f}_-'(t)=2\mathfrak{f}_+(t) \ ,
\end{align}
becomes manifest. As such, the $\mathfrak{f}_-(t)$ can be represented as an integral of $\mathfrak{f}_+(t)$
\begin{align}
\mathfrak{f}_-(t)=-\mathfrak{f}_+(t)-2\int_t^{\infty}\frac{\mathfrak{f}_+(v)}{v}dv \ , 
\end{align}
since both of them decay exponentially at large $t$, which can be seen from a saddle-point analysis of Eq.~(\ref{eq:scalingsimple}). In the appendix.~\ref{sec:cylinder}, by investigating the number observables on the cylinder, 
we can establish alternative integral representations for $\mathfrak{f}_{\pm}(t)$  and further simplify them into products of Bessel-$K$ functions:
\begin{align}
&\mathfrak{f}_+(t)=\frac{t}{2\pi^2}K_0\left(\frac{t}{2}\right)K_1\left(\frac{t}{2}\right)-\frac{t^2}{4\pi^2}\left(K_1^2\left(\frac{t}{2}\right)-K_0^2\left(\frac{t}{2}\right)\right) \ , \label{eq:Besslefp} \\
&\mathfrak{f}_-(t)=\mathfrak{f}_+(t)-\frac{1}{\pi^2}K_0^2\left(\frac{t}{2}\right) \label{eq:Besslefm} \ . 
\end{align}
To show the equivalence to the Barnes representations Eq.~(\ref{eq:scalingsimple}), one can Mellin transform the above with the help of the first Barnes Lemma.

We now show that $\mathfrak{f}_+(t)$ is in fact an integral of the following four-point function in the 2D Ising CFT
\begin{align}
{\cal G}_c(z_1,z_2,z_3,z_4)=\frac{\langle\sigma(z_1)\psi(z_2)\psi(z_3)\sigma(z_4)\rangle-\langle\sigma(z_1)\sigma(z_4)\rangle\langle\psi(z_2)\psi(z_3)\rangle}{\langle \sigma(z_1)\sigma(z_4) \rangle} \ .
\end{align}
 Here, the required configuration is
\begin{align}
z_1=\left(\frac{1}{2},0\right) \ , \  z_4=-\left(\frac{1}{2},0\right) \ , \ z_2=\left(0,\frac{R}{2}+\frac{X}{2}\right)\ , z_3=\left(0,\frac{R}{2}-\frac{X}{2}\right)\ .  \label{eq:config}
\end{align} 
The $z_1$ and $z_4$ correspond to the spin-fields of the two point function separated between the Euclidean time, which we set to $t_E=\pm \frac{1}{2}$. The right-moving fermion fields are placed at the zero Euclidean time, measuring the fermion number. Without losing of generality, we normalize the free right-moving fermion at zero time as
\begin{align}
\psi(x)=\int_0^{\infty}dp \left(a(p)e^{ip\cdot x}+a^{\dagger}(p)e^{-ip\cdot x}\right) \ , \{a(p),a^{\dagger}(p')\}=\delta(p-p') \ . 
\end{align}
As such, the number operator reads
\begin{align}
a^{\dagger}(p)a(p)=\frac{1}{(2\pi)^2}\int dx_2 dx_3 \psi(x_2)\psi(x_3)e^{ip(x_2-x_3)} \ , 
\end{align}
while the equal-time two point function reads
\begin{align}
\langle\psi(x_2)\psi(x_3) \rangle=\frac{1}{-i(x_2-x_3+i0)} \ .
\end{align}
With these normalizations,  the four point function for the configuration Eq.~(\ref{eq:config}) can be computed using the textbook formulas~\cite{DiFrancesco:1997nk} as
\begin{align}
{\cal G}_c\left(z_1,z_2,z_3,z_4\right)=\frac{i}{X+i0} \bigg(\frac{\left(-X^2+R^2+1\right)}{\sqrt{(X-R)^2+1} \sqrt{(X+R)^2+1}}-1\bigg) \ . 
\end{align}
As expected, it is an analytic function in $X$ and $R$, in a neighborhood of the real axis. Given the above, the averaged right-moving fermion number above the momentum $t$ is given by 
\begin{align}
&\tilde f(t)=\frac{1}{2(2\pi)^2}\int_{-\infty}^{\infty} dR \int_{-\infty}^{\infty} dX \frac{e^{it X}}{-i(X+i0) }{\cal G}_c\left(z_1,z_2,z_3,z_4\right) \nonumber \\ 
&=-\frac{1}{2(2\pi)^2}\int_{-\infty}^{\infty} dX e^{it X} \frac{1}{(X+i0)^2}\int_{-\infty}^{\infty} dR\bigg(\frac{\left(-X^2+R^2+1\right)}{\sqrt{(X-R)^2+1} \sqrt{(X+R)^2+1}}-1\bigg) \ . 
\end{align}
Here we show that it is exactly $\mathfrak{f}_+(t)$ . The first step is to perform the $R$ integral, which can be given as 
\begin{align}
-\int_{-\infty}^{\infty} dR\bigg(\frac{\left(-X^2+R^2+1\right)}{\sqrt{(X-R)^2+1} \sqrt{(X+R)^2+1}}-1\bigg)=\pi X^2  \, _2F_1\left(\frac{1}{2},\frac{3}{2};2;-X^2\right) \ .
\end{align}
To establish this identity, one can introduce the Schwinger parameters for the inverse square roots and then Mellin transform the $X$. It is sufficiently regular at $X=0$ such that the $+i0$ is not needed anymore. As such, one has
\begin{align}
\tilde f(t)&=\frac{1}{4\pi }\int_{0}^{\infty} dX \cos tX \, _2F_1\left(\frac{1}{2},\frac{3}{2};2;-X^2\right)\nonumber \\ 
&=\frac{1}{2\pi^2}\int \frac{ds}{2\pi i}\frac{\Gamma\left(\frac{1}{2}+s\right)\Gamma\left(\frac{3}{2}+s\right)\Gamma(-s)}{\Gamma(2+s)} \int_0^{\infty} dX \cos tr(X^2)^{s} \nonumber \\ 
&=\int\frac{dz}{2\pi i}\frac{\Gamma^2\left(\frac{z}{2}\right)\Gamma \left(\frac{z}{2}+1\right)}{8\pi^{\frac{3}{2}}\Gamma\left(\frac{z}{2}+\frac{3}{2}\right)}\left(\frac{t}{2} \right)^{-z} \equiv \mathfrak{f}_+(t)\ . 
\end{align}
 Here, we used the standard Mellin-Barnes representation of the hypergeometric function (this is obtained naturally after the $R$ integral), and changed the variable to $z=2s+1$ in the last step. As such, we have shown that by summing the form factor expansion of a {\it massive} Ising QFT under non-trivial measurements, one reaches an integrated four-point function in the Ising CFT, in the massless limit. In the appendix.~\ref{sec:cylinder}, by doing a more careful calculation with a finite cylinder radius $L$, we can also relate $\mathfrak{f}_-(t)$ to CFT correlators with Ramond fermions.

We now summarize the various important properties of the scaling function, using the representations established above.  First, in the small-$t$ limit, it has double logarithms starting from the quadratic power.  The logarithmic asymptotics is reminiscent of, but much simpler than what we have in the full massive theory. In particular, the number of the logarithms does not blow up and the expansion is absolutely convergent with the $1/(k!)^2$ suppression for the $t^{2k}$ power. At large $t$, on the other hand, from the Bessel function representations Eq.~(\ref{eq:Besslefm}), 
the scaling function decays exponentially 
\begin{align}
\mathfrak{f}_+(t) \rightarrow \frac{e^{-t}}{2\pi t}\bigg(1-\frac{3}{2t}+\frac{33}{8t^2}-\frac{255}{16 t^3}+{\cal O}(t^{-4})\bigg) \ .
\end{align}
It might be helpful to establish spectral representations for the scaling function, which can be achieved by using 
\begin{align}
\, _2F_1\left(\frac{1}{2},\frac{3}{2};2;-X^2\right)=\frac{2}{\pi}\int_0^1 dy \sqrt{\frac{y}{1-y}}\frac{1}{\sqrt{1+yX^2}} \ , 
\end{align}
and Fourier-transform the $X$ to obtain
\begin{align} \label{eq:scaleanter}
\mathfrak{f}_{+}(t)=\frac{1}{4\pi^2}\int_{-\infty}^{\infty}\frac{d\nu}{ \sqrt{\nu^2+1}}\int_0^{1} \frac{dy}{\sqrt{1-y}}\exp \bigg(-t\sqrt{\frac{\nu^2+1}{y}}\bigg) \ . 
\end{align}
From Eq.~(\ref{eq:scaleanter}), it is manifest that the scaling function is the Laplace transform of a tempered distribution supported in $(1,\infty)$ and can be analytically continued to the whole complex plane in $t$, with a branch-cut along the negative real axis. To make contact with the traditional momentum space approach to the semi-inclusive process, it is convenient to recall that we have $t=rE$, and $r=2z=t_E-ix^1 \rightarrow i(t-x^1)=ix^-$. As such,  after analytic continued back to the real time, one has
\begin{align}
\mathfrak{f}_{+}(x^-P^+)=\frac{1}{4\pi^2}\int_{-\infty}^{\infty}\frac{d\nu}{\sqrt{\nu^2+1}}\int_0^{1} \frac{dy}{\sqrt{1-y}}\exp \bigg(-i\sqrt{\frac{\nu^2+1}{y}}x^- P^+\bigg) \ ,
\end{align}
where $E=\frac{e^Y}{2}=P^+$ denotes the ``plus'' momentum threshold. From the above, after taking the $\ln P^+$ derivative and then Fourier-transforming the light-front coordinate $x^-$, one can further obtain the ``fragmentation function'', counting the number of particles with the specific momentum fraction $z=\sqrt{\frac{y}{\nu^2+1}}<1$. The small-$t$ logarithmic singularities of the scaling functions are then encoded by  the small-$z$ properties of the fragmentation function.

\subsection{The small-$r$ expansion of $\Psi_{\pm}(r,Y)$}\label{sec:smallr}
Given the knowledge of the scaling functions and their small $t$ expansions at various powers, we can now find out the full $Y$ dependencies in the small-$r$ expansion of the $V_{\pm}(r, Y)$ and $\Psi_{\pm}(r,Y)$. The crucial information is the Eq.~(\ref{eq:f0minus}) at the leading power, and the Eq.~(\ref{eq:fplus2}), Eq.~(\ref{eq:fminus2}) at ${\cal O}(r^2)$. 

We first consider the leading power. We will be brief here in the argumentation and refer interested reader to the Appendix.~\ref{sec:expansiondetail} for more details.  To start, notice that in the scaling functions, apart from the constants and the $\ln t$ terms, one only has the $t^{1-\sigma}$ term that collapse to ${\cal O}(1)$. The $t^{1-\sigma}$ term must be the lowest term in the $V^{(0)}(r,Y)$, when expanded in $r^{1-\sigma}$. Moreover, this term, when propagates to the high-powers in $r^{1-\sigma}$, will not generate new $Y$-dependencies other than $\sinh (1-\sigma)Y$ . Thus, the $V(r,Y)$ at the leading power must be given by the following form
\begin{align}
V^{(0)}(r,Y)=V_0(r)\sinh (1-\sigma)Y \ , \label{eq:V0}
\end{align}
where boundary condition is given by Eq.~(\ref{eq:f0minus})
\begin{align}
&V_0(r)|_{r\rightarrow 0}=\frac{2^{3 \sigma-3}  \sin \left(\frac{\pi  \sigma}{2}\right) \Gamma \left(\frac{\sigma-1}{2}\right)^2}{\pi ^2}r^{1-\sigma} \ .
\end{align}
The $V_0(r)$ satisfies the following linear equation
\begin{align}
\frac{d^2V_0}{dr^2}+\frac{1}{r}\frac{dV_0}{dr}-\frac{(1-\sigma)^2V_0}{r^2}=\frac{1}{2B^2r^{2\sigma}}\frac{1}{(1+br^{2-2\sigma})^2}V_0(r) \ . 
\end{align}
This is because in $\cosh 2\varphi$, the dominant term when $\sigma \rightarrow 1$ is exactly $\frac{1}{2B^2r^{2\sigma}(1+br^{2-2\sigma})^2}$, while all the corrections are ${\cal O}(r^4)$ (with respect to the dominant term). This equation is of the type Eq.~(\ref{eq:diffgen}) and is solved by 
\begin{align}
V_0(r)=r^{1-\sigma}\frac{2^{3 \sigma-3}  \sin \left(\frac{\pi  \sigma}{2}\right) \Gamma \left(\frac{\sigma-1}{2}\right)^2}{\pi ^2}\frac{16B^2(1-\sigma)^2}{16B^2(1-\sigma)^2-r^{2-2\sigma}} \ . \label{eq:solutionV0}
\end{align}
To obtain the $\Psi(r,Y)$ at this power, we also need to include the leading-power part of the first term in the decomposition Eq.~(\ref{eq:decompPsi}). By taking the $\sigma$ derivative of $\varphi^{(0)}$ in Eq.~(\ref{eq:phi0}), this term can also be determined as 
\begin{align}
\frac{\lambda}{2}\partial_{\lambda} \varphi\bigg|^{(0)}_{\ln r} =-\frac{\sin\frac{\pi\sigma}{2}}{\pi\cos \frac{\pi \sigma}{2}}\bigg(\frac{16B^2(1-\sigma)^2+r^{2-2\sigma}}{16B^2(1-\sigma)^2-r^{2-2\sigma}}\bigg) \ . 
\end{align}
As such, the $\Psi^{(0)}(r,Y)$ can be determined as 
\begin{align}
&\Psi^{(0)}(r,Y)=-\frac{\sin\frac{\pi\sigma}{2}}{\pi\cos \frac{\pi \sigma}{2}}\bigg(\frac{16B^2(1-\sigma)^2+r^{2-2\sigma}}{16B^2(1-\sigma)^2-r^{2-2\sigma}}\bigg) Y\nonumber \\ 
&+r^{1-\sigma}\frac{2^{3 \sigma-3}  \sin \left(\frac{\pi  \sigma}{2}\right) \Gamma \left(\frac{\sigma-1}{2}\right)^2}{\pi ^2}\frac{16B^2(1-\sigma)^2}{16B^2(1-\sigma)^2-r^{2-2\sigma}}\sinh(1-\sigma)Y \ .
\end{align}
In the $\sigma \rightarrow 1$ limit, it is finite 
\begin{align}
\lim_{\sigma\rightarrow 1} \Psi^{(0)}(r,Y)=\frac{Y^3}{3\pi^2}\frac{1}{\Omega}-\frac{Y\Omega}{\pi^2} \ . 
\end{align}
Again, the finiteness is achieved only after the singularity cancellation between the two terms. To determine the observable $N_{\pm}(r,Y)$, one also needs to determine the leading-power part of $\Xi=\Psi_++\Psi_-$. One clearly has
\begin{align}
\bigg(\frac{d^2}{dr^2}+\frac{1}{r}\frac{d}{dr}-\frac{1}{r^2}\frac{\partial^2}{\partial Y^2}\bigg)\Xi^{(0)}(r,Y)=-\frac{1}{2B^2r^{2\sigma}}\frac{1}{(1+br^{2-2\sigma})^2}\Psi^{(0)}(r,Y) \ .
\end{align}
Since in the scaling function $f_{+}(t)$ for $\Xi+\Psi=2\Psi_+$, there are no $t^{1-\sigma}$ terms at all, one must has
\begin{align}
\Xi^{(0)}(r,Y)=-\Psi^{(0)}(r,Y)-\lambda\frac{d}{d\lambda}\bigg(\sigma_{+}\bigg) Y =-\Psi^{(0)}(r,Y)-\frac{(1-\sigma)\tan \frac{\pi \sigma}{2}}{\pi}Y \ .
\end{align}
It is also finite in the $\sigma \rightarrow 1$ limit
\begin{align}
\lim_{\sigma\rightarrow 1} \Xi^{(0)}(r,Y)=-\frac{Y^3}{3\pi^2}\frac{1}{\Omega}+\frac{Y\Omega}{\pi^2}-\frac{2}{\pi^2}Y \ . 
\end{align}
The above can be used to determine the $N_{\pm}(r,Y)$ up to the next-to-leading power
\begin{align}
&N_{\pm}(r,Y;\sigma=1)=\frac{1}{\pi^2}\left(\Omega-1-Y\right) \pm r\bigg(\frac{-2\Omega^3+\zeta_3-Y^3+3Y\Omega^2}{6\pi^2}\bigg)  \  .
\end{align}
The combination in the power correction is interesting. We can write
\begin{align}
\frac{-2\Omega^3-Y^3+3Y\Omega^2+\zeta_3}{6\pi^2}=\frac{-(\Omega-Y)^2(Y+2\Omega)+\zeta_3}{6\pi^2}=-\frac{L^2\Omega}{2\pi^2}+\frac{L^3+\zeta_3}{6\pi^2} \ , 
\end{align}
where we have introduced the logarithm $L=\Omega-Y$ for the scaling variable again. As such, the leading power is just the $\mathfrak{f}_+^{0)}(t)$, while at the next-to-leading power there are two terms. The first term is proportional to the $-\eta(r) (\mathfrak{f}_+^{(0)}(t)-\mathfrak{f}_-^{(0)}(t))$ and can still be expressed by the scaling functions $\mathfrak{f}_{\pm}^{(0)}$, while the second term $L^3$ is due to the scaling-violating effects in the $V(r,Y)$. We will see that this pattern continues to hold at the high-powers. 

We now move to the ${\cal O}(r^2)$ order. From Eq.~(\ref{eq:fplus2}), Eq.~(\ref{eq:fminus2}), the small $t$ limits of the scaling function is given by
\begin{align}
f_{+}^{(2)}(t)-f_{-}^{(2)}(t)=f_1(\sigma)t^2+f_2(\sigma)t^{\sigma+1}+f_3(\sigma)t^{3-\sigma} \ ,
\end{align}
with 
\begin{align}
&f_1(\sigma)=\frac{\tan \left(\frac{\pi  \sigma}{2}\right)}{4 \pi  (\sigma^2-1)} \ , \\ 
&f_2(\sigma)=\frac{2^{-\sigma-1} \sin \left(\frac{\pi  \sigma}{2}\right) \Gamma \left(\frac{1}{2} (-\sigma-1)\right) \Gamma (-\sigma-1)}{\pi ^{3/2} \Gamma \left(-\frac{\sigma}{2}\right)} \ , \\
&f_3(\sigma)=-\frac{2^{2 \sigma-7} (\sigma-2) \sin \left(\frac{\pi  \sigma}{2}\right) \Gamma \left(\frac{\sigma-3}{2}\right)^2}{\pi ^2} \ . 
\end{align}
Clearly, since in the $\cosh 2\varphi$ there are no quadratic power corrections with respect to the dominant term $\frac{1}{2B^2r^{2\sigma}(1+br^{2-2\sigma})^2}$, the solution
$V^{(0)}(r,Y)$ in Eq.~(\ref{eq:V0}) will not propagate to this power. As such, we can write
\begin{align}
V^{(2)}(r,Y)=F_1(r)r^2\sinh 2Y+F_2(r)r^{\sigma+1}\sinh (\sigma+1)Y+F_3(r)r^{3-\sigma}\sinh(3-\sigma)Y \ . 
\end{align}
They satisfy the equations
\begin{align}
&\bigg(\frac{d^2}{dr^2}+\frac{1}{r}\frac{d}{dr}-\frac{2^2}{r^2}\bigg)F_1(r)r^2=\frac{1}{2B^2 r^{2\sigma}(1+br^{2-2\sigma})^2}F_1(r)r^2 \ , \\
&\bigg(\frac{d^2}{dr^2}+\frac{1}{r}\frac{d}{dr}-\frac{(1+\sigma)^2}{r^2}\bigg)F_2(r)r^{1+\sigma}=\frac{1}{2B^2 r^{2\sigma}(1+br^{2-2\sigma})^2}F_2(r)r^{1+\sigma}  \ , \\ 
&\bigg(\frac{d^2}{dr^2}+\frac{1}{r}\frac{d}{dr}-\frac{(3-\sigma)^2}{r^2}\bigg)F_3(r)r^{3-\sigma}=\frac{1}{2B^2 r^{2\sigma}(1+br^{2-2\sigma})^2}F_3(r)r^{3-\sigma} \ . 
\end{align}
They are of the type Eq.~(\ref{eq:diffgen}) again and are solved by 
\begin{align}
&F_1(r)=2^{1-2}f_1(\sigma)\bigg(1+\frac{-2br^{2-2\sigma}}{1+br^{2-2\sigma}}\frac{1-\sigma}{3-\sigma}\bigg) \ , \\
&F_2(r)=2^{1-(\sigma+1)}f_2(\sigma)\bigg(1+\frac{-2br^{2-2\sigma}}{1+br^{2-2\sigma}}\frac{1-\sigma}{2}\bigg) \ , \\
&F_3(r)=2^{1-(3-\sigma)}f_3(\sigma)\bigg(1+\frac{-2br^{2-2\sigma}}{1+br^{2-2\sigma}}\frac{1-\sigma}{4-2\sigma}\bigg) \ . 
\end{align}
For the $Z(r,Y)$, if we write the scaling function as 
\begin{align}
f_{+}^{(2)}(t)+f_{-}^{(2)}(t)=\tilde f_1(\sigma)t^2+\tilde f_2(\sigma)t^{\sigma+1}+\tilde f_3(\sigma)t^{3-\sigma} \ ,
\end{align}
then one can show that $Z(r,Y)$ is solved by 
\begin{align}
Z^{(2)}(r,Y)=\tilde F_1(r)r^2\sinh 2Y+\tilde F_2(r)r^{\sigma+1}\sinh (\sigma+1)Y+\tilde F_3(r)r^{3-\sigma}\sinh(3-\sigma)Y \ , 
\end{align}
with 
\begin{align}
\tilde F_1(r)=2^{1-2}(\tilde f_1+f_1)(\sigma)-F_1(r) \ , \\
\tilde F_2(r)=2^{1-(\sigma+1)}(\tilde f_2+f_2)(\sigma)-F_2(r) \ , \\
\tilde F_3(r)=2^{1-(3-\sigma)}(\tilde f_3+f_3)(\sigma)-F_3(r) \ . 
\end{align}
From the above,  after taking the $\sigma \rightarrow 1$ limit, one obtains
\begin{align}
V^{(2)}(r,Y;\sigma=1)=\frac{(L+1)((2 \Omega+1)L+1)}{16 \pi ^2 \Omega}t^2-(Y\rightarrow -Y) \ . 
\end{align}
Here $L=\Omega-Y$ and $t=\frac{r}{2}e^{Y}$. As such, the leading power part for $V$ in the scaling limit does not scale. It can be written as
\begin{align}
V^{(2)}(r,Y;\sigma=1)=t^2\frac{L(L+1)}{8\pi^2}{\color{red}+t^2\frac{(L+1)^2}{16\pi^2\Omega}}-(Y\rightarrow -Y) \ .  \label{eq:Vtwo}
\end{align}
The first term is just given by the scaling function , while the scaling violating term shown in red is again suppressed by $\frac{1}{\Omega}$. On the other hand, for $V_+(r,Y)$, one has
\begin{align}
V_+^{(2)}(r,Y)=f_+^{(2)}\left(\frac{r}{2}e^Y\right)-f_+^{(2)}\left(\frac{r}{2}e^{-Y}\right) \ . \label{eq:Vplustwo}
\end{align}
It is again expressed by the scaling function $f_+$. Also notice that at the quadratic order $V^{(2)}=\Psi^{(2)}={\cal Q}^{(2)}$, since the $\lambda\partial_\lambda \phi_{\pm}$ contains no quadratic power corrections.  The above can be used to determine the corrections to $N_{\pm}(r,Y)$ up to ${\cal O}(r^3)$ in the small-$r$ limit. The results read
\begin{align}
&N_{\pm}(r,Y;\sigma=1)=\frac{1}{2}N_{\pm}(r)-\frac{Y}{\pi^2}+V_+^{(2)}(r,Y;\sigma=1)\nonumber \\ 
&\mp\frac{r\Omega}{2\pm r\Omega}\bigg(\frac{Y^3}{3\pi^2}\frac{1}{\Omega}-\frac{Y\Omega}{\pi^2}+V^{(2)}(r,Y;\sigma=1)\bigg)+{\cal O}(r^4) \ . \label{eq:expandNsmallr}
\end{align}
Here, $N_{\pm}(r)$ are the total number observables given by Eq.~(\ref{eq:expandNpm}), and the $V^{(2)}$, $V_+^{(2)}$ are given by the equations Eq.~(\ref{eq:Vtwo}), Eq.~(\ref{eq:Vplustwo}). We can express the $Y$ and $\Omega$ dependencies explicitly. For the $\Psi_+$, one has 
\begin{align}
&\Psi_+^{(2)}(r,Y;\sigma=1)=V_+^{(2)}(r,Y;\sigma=1)\equiv\frac{r^2}{32\pi^2}\sum_{k=0}^{2}\psi_k(Y)\Omega^k \ ,  \label{eq:Psiplus2expli}\\ 
&\psi_0(Y)=-4 Y \cosh 2Y+ 
   (3 + 2 Y^2) \sinh 2Y,  \ \\ 
   &\psi_1(Y)=4(-Y\cosh 2Y+ \sinh 2Y) \  , \  \psi_2(Y)= 2\sinh 2Y \ . 
\end{align}
Similarly, for $\Psi$ one has
\begin{align}
&\Psi^{(2)}(r,Y;\sigma=1)=V^{(2)}(r,Y;\sigma=1)=\frac{r^2}{32\pi^2}\sum_{k=-1}^{2}\varphi_k(Y)\Omega^{k} \ , \label{eq:Psi2expli} \\
& \varphi_{-1}(Y)=-2 Y \cosh 2Y + (1 + Y^2) \sinh 2Y \ , \\
& \varphi_{0}(Y)=2 \left(Y^2+1\right) \sinh 2 Y-4 Y \cosh 2 Y \ , \\
&\varphi_1(Y)=3 \sinh 2 Y-4 Y \cosh 2 Y \ , \ \varphi_2(Y)=2 \sinh 2 Y \ . 
\end{align}
Here we emphasize that the Eq.~(\ref{eq:expandNsmallr}) is the small-$r$ expansion of the observables at an arbitrary $Y$. Also notice that from the Eq.~(\ref{eq:boost}), the $Y$ dependency can be combined with the phase angle of the $z$ as
$z \rightarrow\frac{r}{2}e^{Y-i\vartheta}$, $\bar z \rightarrow\frac{r}{2}e^{-Y+i\vartheta}$. As such, as far as $|\vartheta|<\frac{\pi}{2}$, the $Y$ dependency can be analytic continued to obtain the angle dependency of the observables as well. Using our convention for the momentum $2p=x-x^{-1}$, one has $t_E=r\cos\vartheta$, $x^1=r\sin \vartheta$. As such, the angel dependent observable probes the number of particles above the rapidity threshold $Y$, underlying the form factor expansion of an Euclidean two-point function with the direction vector $(t_E,x^1)=(r\cos \vartheta,r\sin\vartheta)$. In particular, for $Y=0$, it measures the number of particles whose momentum has an angle $\frac{\pi}{2}-\vartheta$ relative to the direction vector.

\subsection{The scaling limit at $\sigma=1$}\label{sec:scalinglimit}
We have shown before that for $0<\sigma<1$, the naive scaling limit of the exponential form factor expansion leads to the scaling functions that capture the largest contribution in $Y$ at each power in $r$. This allows to determine the integration constants in $V_{\pm}(r,Y)$ and obtain the full-$Y$ dependency at each power in $r$. On the other hand, it is clear from the explicit formulas below the Eq.~(\ref{eq:Psiplus2expli}), Eq.~(\ref{eq:Psi2expli}), that the small-$r$ expansion at $\sigma=1$ also suffers increasing powers of $e^{Y}$, and in the scaling limit these enhancements must be resumed. It is interesting to see if the scaling functions $f_{\pm}(t)$ continue to play a role at $\sigma=1$. This is not clear at the first glance. The point is that although $f_{\pm}(t)$ themselves are finite at $\sigma=1$, there are other contributions which are power suppressed when $0<\sigma<1$, but also collapse to the same power described by the $f_{\pm}(t)$.  Notice that in the scaling limit, $t$ is fixed to ${\cal O}(1)$ and one keeps track the powers in remaining variable $r$. At each power, the contributions can depend on $t$ arbitrarily, while on $r$ logarithmically. The task now is to find out explicitly the
contributions in ${\cal Q}$ and ${\cal H}$ at the leading power in the scaling limit.

The results up to $r^2$ in the previous subsection already demonstrate these points. We have 
\begin{align}
&{\cal Q}^{(0)}(r,Y;\sigma=1)=\frac{2\Omega^3-\zeta_3+Y^3-3Y\Omega}{3\pi^2\Omega}=\frac{L^2}{\pi^2}{\color{red}-\frac{L^3+\zeta_3}{3\pi^2 \Omega}} \ , \label{eq:Q0} \\  
&{\cal Q}^{(2)}(r,Y;\sigma=1)=t^2\frac{L(L+1)}{8\pi^2}{\color{red}+t^2\frac{(L+1)^2}{16\pi^2\Omega}}-(Y\rightarrow -Y) \ . \label{eq:Q1}
\end{align}
Notice that the logarithm $L$ in the scaling variable $t$ is defined in Eq.~(\ref{eq:defL}). As such, there are indeed scaling-violating terms of ${\cal Q}$ at the leading power as shown in red, but up to ${\cal O}(t^2)$ they depend on $r$ only through $\frac{1}{\Omega}$. Not only that, we notice that 
\begin{align}
&t\frac{d}{dt} \bigg(-\frac{L^3+\zeta_3}{3\pi^2 }\bigg)=\frac{L^2}{\pi^2} \ , \\
&t\frac{d}{dt} \bigg(t^2\frac{(L+1)^2}{16\pi^2}\bigg)=t^2\frac{L(L+1)}{8\pi^2} \ . 
\end{align}
Thus, the above strongly suggest that at the leading power in the scaling limit, we have 
\begin{align}
{\cal Q}^{(0)}(t,r;\sigma=1)=\mathfrak{f}_+(t)-\mathfrak{f}_-(t)+\frac{\mathfrak{g}(t)}{\Omega} \ , \label{eq:QLeading1} \\
t\frac{d\mathfrak{g}(t)}{dt}=\mathfrak{f}_+(t)-\mathfrak{f}_-(t) \ , \ \mathfrak{g}(t)|_{t\rightarrow 0}=-\frac{L^3+\zeta_3}{3\pi^2}+{\cal O}(t^2) \ . \label{eq:QLeading2}
\end{align}
Let's show that this is indeed the case to all orders in $t$. For this purpose, we change the variable from $(r,Y)$ to $(t,r)$. Then, acting on any function $f(t,r)$, one has 
\begin{align}
\bigg(\frac{\partial^2}{\partial r^2}+\frac{1}{r}\frac{\partial}{\partial r}-\frac{1}{r^2}\frac{\partial^2}{\partial Y^2}\bigg)f(r,Y)=\bigg(\frac{\partial^2}{\partial r^2}+\frac{1}{r}\frac{\partial}{\partial r}+\frac{2t}{r}\frac{\partial^2}{\partial r \partial t}\bigg)f(t,r) \ . \label{eq:laplacetr} 
\end{align}
This equation allow the homogeneous solution at any power $r^{b}$ with $b>0$
\begin{align}
r^b t^{-\frac{b}{2}}=2^{\frac{b}{2}} (r \exp(-Y))^{\frac{b}{2}} \ , \label{eq:homogentr}
\end{align}
On the other hand, at $b=0$, it allows an arbitrary function in $t$. Clearly, this arbitrary function exactly corresponds to the scaling function that picks up the largest $e^Y$ terms at each order in $r$. On the other hand, the homogeneous solution probes the largest $e^{-Y}$ terms, which are opposite to the largest $e^{Y}$ terms due to the $Y$-parity of the $\Psi_{\pm}(r,Y)$.  As such, the knowledge of the scaling function also allows to determine the power expansion in $r$ at fixed $t$. Here we show how it works. At the leading power, we can write 
\begin{align}
\frac{\lambda}{2} \partial_\lambda \phi^{(0)}=\frac{\sin \frac{\pi \sigma}{2}}{\pi\cos\frac{\pi\sigma}{2}}\bigg(\frac{br^{2-2\sigma}-1}{1+br^{2-2\sigma}}\ln r+\frac{B'}{B}\frac{br^{2-2\sigma}-1}{1+br^{2-2\sigma}}-\frac{2}{1-\sigma}\frac{b r^{2-2\sigma}}{1+br^{2-2\sigma}}\bigg) \ .
\end{align}
The above implies that at the leading power one can write
\begin{align}
&{\cal Q}^{(0)}(t,r)=(f_+-f_-)(t)+\frac{\sin \frac{\pi \sigma}{2}}{\pi \cos \frac{\pi \sigma}{2}}\left(-\frac{2br^{2-2\sigma}}{1+br^{2-2\sigma}}\right)\bigg(\ln \frac{1}{2t}-\frac{B'}{B}+\frac{1}{1-\sigma}\bigg)+ \  ... \ ,
\end{align}
which suggests the following general form at the leading small-$r$ power  
\begin{align}
{\cal Q}^{(0)}(t,r)=(f_+-f_-)(t)-\frac{2b(1-\sigma)r^{2-2\sigma}}{1+br^{2-2\sigma}}g(t) \ .\label{eq:QtrLead}
\end{align}
Plug in the differential equation Eq.~(\ref{eq:lineardifQ}), keeping only the dominant contribution in the $\cosh 2\phi$, and using the Eq.~(\ref{eq:laplacetr}), the $r^{2-2\sigma}$ dependencies drop, and one ends up at a closed first order equation for $g(t)$
\begin{align}
tg'(t)+(1-\sigma)g(t)=(f_+-f_-)(t)\ . \label{eq:differg}
\end{align}
The only homogeneous solution is given by $t^{\sigma-1}$. It corresponds to the $e^{-(1-\sigma)Y}$ in the $\sinh(1-\sigma)Y$ term and is given by $f_-^{(0)}$ in Eq.~(\ref{eq:f0minus})
\begin{align}
&-r^{1-\sigma}\frac{2^{3 \sigma-4}  \sin \left(\frac{\pi  \sigma}{2}\right) \Gamma \left(\frac{\sigma-1}{2}\right)^2}{\pi ^2}\frac{1}{1+br^{2-2\sigma}}e^{-(1-\sigma)Y}\nonumber \\ 
&=-t^{\sigma-1}\frac{2^{4 \sigma-5}  \sin \left(\frac{\pi  \sigma}{2}\right) \Gamma \left(\frac{\sigma-1}{2}\right)^2}{\pi ^2}\frac{r^{2-2\sigma}}{1+br^{2-2\sigma}} \ .
\end{align}
This is consistent with the fact that the only $e^{-Y}$ completion of the scaling function that remains at the leading power as $\sigma \rightarrow 1$ is due to the $t^{1-\sigma}$ term, corresponding to the integration constant of Eq.~(\ref{eq:differg}). The solution to $g(t)$ is then given by
\begin{align}
g(t)=\frac{t^{\sigma-1}}{2b}\frac{2^{4 \sigma-5}  \sin \left(\frac{\pi  \sigma}{2}\right) \Gamma \left(\frac{\sigma-1}{2}\right)^2}{\pi ^2(1-\sigma)}+t^{\sigma-1}\int_0^{t}v^{-\sigma}(f_+(v)-f_-(v))dv \ . \label{eq:solug}
\end{align}
Clearly, for the small-$v$ expansion of $f_+-f_-$ beyond the first order, the integral is convergent at $\sigma=1$, while the pere-factor is also finite 
\begin{align}
-\frac{2(1-\sigma)br^{2-2\sigma}}{1+br^{2-2\sigma}} \rightarrow \frac{1}{\Omega} \ . 
\end{align}
On the other hand, the integration in Eq.~(\ref{eq:solug}) of $f_+^{(0)}-f_-^{(0)}$ is divergent when $\sigma \rightarrow 1$, but the singularities are exactly canceled by the first term. Using 
\begin{align}
f_+^{(0)}(t)-f_-^{(0)}(t)=\frac{\sin \frac{\pi \sigma}{2}}{\pi\cos\frac{\pi\sigma}{2}}\left(\ln \frac{1}{2t}-\frac{B'}{B}\right)+\frac{2^{2 \sigma-3} t^{1-\sigma} \sin \left(\frac{\pi  \sigma}{2}\right) \Gamma \left(\frac{\sigma-1}{2}\right)^2}{\pi ^2} \ , 
\end{align}
it is not hard to see that in the $\sigma \rightarrow1$ limit, one has
\begin{align}
\frac{t^{\sigma-1}}{2b}\frac{2^{4 \sigma-5}  \sin \left(\frac{\pi  \sigma}{2}\right) \Gamma \left(\frac{\sigma-1}{2}\right)^2}{\pi ^2(1-\sigma)}+t^{\sigma-1}\int_0^{t}v^{-\sigma}(f^{(0)}_+(v)-f^{(0)}_-(v))dv\rightarrow -\frac{L^3+\zeta_3}{3\pi^2} \ ,  
\end{align}
which agrees exactly with Eq.~(\ref{eq:Q0}). As such, the relations Eq.~(\ref{eq:QLeading1}) and Eq.~(\ref{eq:QLeading2}) that resum the scaling-violating effects are therefore established. Given the ${\cal Q}(t,r)$, we can determine the ${\cal H}(t,r)$ at the leading power as
\begin{align}
{\cal H}^{(0)}(t,r)=(f_++f_-)(t)+\frac{2b(1-\sigma)r^{2-2\sigma}}{1+br^{2-2\sigma}}g(t) \ .
\end{align}
Clearly, this is because that the $f_+(t)$ contains no exponent $b$ for which the homogeneous solutions Eq.~(\ref{eq:homogentr}) could collapse to ${\cal O}(1)$ when $\sigma \rightarrow 1$ .  

After finding out the leading power part in the scaling limit, here we comment on the structure of the power corrections. A crucial difference to the small-$r$ expansion at finite $Y$ is that we have no quadratic power corrections to ${\cal Q}(t,r)$ and ${\cal H}(t,r)$. Indeed, to promote a term $t^{2\alpha}$ in the scaling function to the full solution of the homogeneous equation, we always need to add the $-e^{-Y}$ part to make the $\Psi_{\pm}$ antisymmetric. The resulting contribution $t^{2\alpha}-\frac{r^{4\alpha}}{t^{2\alpha}2^{4\alpha}}$ will either be at the leading power or at $4\alpha$ power higher, when $\alpha >0$. After multiplying with the dominant term in $\cosh 2\phi$ in the Eq.~(\ref{eq:lineardifQ}), the largest contributions in the generated solution are exactly resumed by Eq.~(\ref{eq:QtrLead}). This re-summation has taken care of all the contributions for which $\alpha \rightarrow 0$ when $\sigma \rightarrow 1$, as well as all the $t^{2\alpha}$ terms for which $\alpha \rightarrow 1$ or higher. As such, corrections to Eq.~(\ref{eq:QtrLead}) are due to the 
neglected $\frac{r^{4\alpha}}{t^{2\alpha}2^{4\alpha}}$ terms, as well as the corrections to the dominant term in $\cosh 2\phi $, which are all at least quartic. Furthermore, they can be computed iteratively power by power in $r$, using the differential equations to resum the attached functions in $r^{2-2\sigma}$. For example, to obtain the ${\cal O}(r^4)$ corrections,  there are two types of contributions. The first is the $e^{-2Y}$ parts associated to $f_{\pm}^{(2)}(t)$ in Eq.~(\ref{eq:fplus2}), Eq.~(\ref{eq:fminus2}). They will be attached to functions in $r^{2-2\sigma}$ after multiplying the dominant term in $\cosh 2\phi$. The second is the Eq.~(\ref{eq:QtrLead}) propagated to the quartic power, after multiplying the subleading contributions in $\cosh 2\phi$.  

To summarize, the results obtained so far allow to obtain the corrections to the observables $N_{\pm}(t,r)$ in the scaling limit $r\rightarrow 0$, $t=\frac{r}{2}e^Y={\cal O}(1)$, up to ${\cal O}(r^3)$ 
\begin{align}
N_{\pm}(t,r;\sigma=1)=\mathfrak{f}_{+}(t)\mp\frac{r}{2\pm r\Omega} \bigg( \left(\mathfrak{f}_+(t)-\mathfrak{f}_-(t)\right)\Omega+\mathfrak{g}(t)\bigg)+{\cal O}(r^4) \ .  \label{eq:finalasymptotics}
\end{align}
As a reminder, the logarithm $\Omega$ is defined in Eq.~(\ref{eq:defomega}), the scaling functions $\mathfrak{f}_{\pm}(t)$ are given by Eq.~(\ref{eq:scalingsimple}), and $\mathfrak{g}(t)$ relates to $\mathfrak{f}_{\pm}(t)$ through Eq.~(\ref{eq:QLeading2}). It is not hard to show that $\mathfrak{g}(t)$ is given by the following Barnes integral with ${\rm Re}(z)>0$
\begin{align}
\mathfrak{g}(t)=-\int \frac{dz}{2\pi i }t^{-z}\frac{2^{z} (z+1) \Gamma \left(\frac{z}{2}\right)^3}{8\pi ^{3/2} z \Gamma \left(\frac{z+3}{2}\right) } \ , 
\end{align}
which exactly generates the leading small-$t$ behavior in Eq.~(\ref{eq:QLeading2}).  Comparing with the small-$r$ expansion at a fixed $Y$, the expansion in the scaling limit turns out considerably simpler at the end. One can further introduce the angle dependency by the analytic continuation $t\rightarrow te^{-i\vartheta}$. As such, we have finally shown that the scaling functions continue to be essential in the scaling limit at $\sigma=1$. In the Appendix.~\ref{sec:numer}, we compare the asymptotic formula Eq.~(\ref{eq:finalasymptotics}) numerically with truncated form factor expansion to further support its correctness.

\section{Conclusion and comments}\label{sec:conclude}
In this work, we have shown that after weighting the form factor expansions of the spin-spin Euclidean two-point correlators in the free-fermion Ising QFT, the weighted two-point functions, measuring the fermion number above a low-rapidity threshold, continue to be controlled by the Sinh-Gordon-type differential equations and allow the exact analysis of the short distance asymptotics. Two scaling functions are crucial in our analysis. They resum the largest power in $e^Y$ at each power in the small-$r$ expansion when $0<\sigma<1$, providing the missing constants that can not be fixed by the differential equations. The computation of the scaling functions is based on the properties of an integral operator and generalize the computations in the traditional Ising connecting problem. At $\sigma=1$, the scaling functions are given by integrated four-point functions in the Ising CFT and dominate the asymptotics of the fermion number observables up to ${\cal O}(r^3)$. 

Before ending the paper, we would like to make the following comments. 
\begin{enumerate}
    \item Although in this work we only consider a specific number observable, the construction in principle can be generalized to more general one-particle measurements. In particular, if we consider the following observable
  \begin{align}
  O(n,p(\theta_i))=\sum_{i=1}^np(\theta_i) \ , 
  \end{align}
with a general measurement function $p(\theta)$ that can depend on extra parameters, then by introducing the 
\begin{align}
E_p=\exp \left(\frac{1}{2}\alpha p(\theta)\right) \ , 
\end{align}
in the Eq.~(\ref{eq:defKgeneral}) and taking the $\alpha$ derivative, most of the constructions can be generalized. Taking second-order derivative in $\alpha$, one can further obtain linear differential equations for two-particle measurements $\sum_{i,j}p(\theta_i)p(\theta_j)$ sourced by the first moment $\sum_{i}p(\theta_i)$.  On the other hand, it is not always possible to transmute the parameter dependencies in $p(\theta)$ into $\bar z,z$, and in general one needs the $\partial,\bar \partial$ derivatives in the differential equations, on top of the parameters in $p$.
\item The $N_{\pm}(r)$ in Eq.~(\ref{eq:expandNpm}), counting the averaged number of particles without constraints, are IR-sensitive even at the leading power. Indeed, after restoring the fermion-mass $m$ dependency, the $\Omega$ contains an explicit $\ln m$ that does not belong to the Ising CFT. The $N_{\pm}(t,r)$ with the lower rapidity threshold, on the other hand, are IR-safe at the leading power in the scaling limit. Indeed, in the Eq.~(\ref{eq:finalasymptotics}), the scaling variable $t$ can be written as 
\begin{align}
t=\frac{mr}{2}e^Y=\frac{r(E+P)}{2}\rightarrow rE \ , 
\end{align}
where $E=m\cosh Y,P=m\sinh Y$ denote the energy and momentum at the lower rapidity threshold $Y$. In the scaling limit, both $r$, $E$, $P$ are UV scales. As such, the scaling function $f_{+}(t)$ depends only on the ratio of two UV scales, indicating the IR safety. This is consistent with the fact that $\mathfrak{f}_+(t)$ can be expressed as a convergent integral of a four-point function in the Ising CFT. Furthermore, the coefficient $\frac{1}{\pi^2}$ of the logarithmic term in $\frac{1}{2}N_{\pm}(r)$ is  the same as that of the $\mathfrak{f}_+^{(0)}(t)$ in Eq.~(\ref{eq:fplus0pi}) and continues to be of CFT origin.  
\item The fact that the fermion number observables can be reduced to Euclidean four-point functions and allow CFT-driven massless limits with IR safety, although true in the free-fermion Ising QFT, can not be expected to be a general property of the Ising universality class. Instead, it relies on the following facts: within all the Ising QFTs~\cite{Fonseca:2006au,Zamolodchikov:2013ama} sharing the common UV fixed point, it is the free-fermion QFT that has a conserved fermion number, as well as the shortest conceptual distance between the massive asymptotic states in the IR, and the massless fermions in the UV. It would be a big surprise to the author, that the integrated four-point function, a well-defined quantity in the Ising CFT with fantastic logarithmic asymptotics, could continue to dominate the particle number observables at small distances, in Ising QFTs without the on-shell freeness. 
\item Nevertheless, it is still not truly trivial to see that the massless limit of a four-point function could be reached, in a way that transpasses the conceptual distance between the spin operator and the free fermions--a gap wider than what could be probed in most NLSM-type large-$N$ examples, by summing the weighted form-factors with massive particles in the asymptotic states. What one gets from the non-Gaussianity are the following: as the virtualities injected to the operators increase, the operators tend to dissipate the high virtuality to a large cloud of ``soft'' particles, whose energies are logarithmicaly suppressed.  The situation is then a bit similar to the $\langle\bar\psi \psi(x)\bar\psi\psi(0)\rangle$-type correlators in the massless Schwinger model. On the other hand, the  particle number scaling functions in that case are given by the trivial one-particle expressions. In the massive fermion Ising, there are more correlations in the created fermions, and the scaling functions are also less simple. But not too much: at the end, the scaling functions, quantities that could be defined in purely massless setups, just happened to coincide with the two-particle form factor integrals, but in the original massive theory (a fact we show in the appendix.~\ref{sec:cylinder}).

\end{enumerate}

\appendix

\section{The number observables of the Ising CFT on a cylinder}\label{sec:cylinder}
In this appendix, we consider the number observables in the Ising CFT, on a finite cylinder with a radius $L$. These considerations in fact lead to major simplifications of the scaling functions $\mathfrak{f}_{\pm}(t)$.  The coordinate on the cylinder reads
\begin{align}
z=e^{\frac{2\pi i}{L}(x+iy)} \ , 
\end{align}
where $y$ is the imaginary time direction, and $0<x<L$ is the spatial direction. We locate the spin operators at 
\begin{align}
z_1=e^{-\frac{2\pi r}{L}}\equiv \alpha<1 \ ,z_4= e^{\frac{2\pi r}{L}}\equiv \alpha^{-1} \ ,
\end{align}
and the fermion operators at
\begin{align}
z_2=e^{\frac{2\pi i}{L}x_2},  \ z_3=e^{\frac{2\pi i}{L}x_3} \ . 
\end{align}
We chose $|z_1|<|z_2|=1-\epsilon<|z_3|=1+\epsilon<z_4$. On the cylinder, there are normalized ``ground states'' $|\Omega_{NS}\rangle$, $|\Omega_R^+\rangle$ and $|\Omega_R^-\rangle$. Only the $|\Omega_{NS}\rangle$ (charge even) and $|\Omega_R^-\rangle$ (charge odd) are consistent with the charge condition. The ground-state energy difference $E_R-E_{NS}=\frac{\pi }{4L}$ approaches $0$ as $L\rightarrow \infty$.

The purpose of this appendix is to compute the fermion-number averages
\begin{align}
&N_{NS}\left(t=\frac{4\pi r }{L}n_0,\alpha\right)=\frac{\langle \Omega_{NS}|\sigma(z_1)\hat N_R(n_0)\sigma(z_4)|\Omega_{NS} \rangle}{\langle \Omega_{NS}|\sigma(z_1) \sigma(z_4)|\Omega_{NS}\rangle} \ , \\
&N_{R}\left(t=\frac{2\pi r }{L}(2n_0+1),\alpha\right)=\frac{\langle \Omega_{R}^-|\sigma(z_1)\hat N_{NS}(n_0)\sigma(z_4)|\Omega^-_{R} \rangle}{\langle \Omega_{R}^-|\sigma(z_1) \sigma(z_4)|\Omega_{R}^-\rangle} \ ,
\end{align}
to show how the scaling functions $\mathfrak{f}_{\pm}(t)$ arise from their large $L$ ($\alpha\rightarrow 1$) limits. The fermion-number observables $\hat N_{R}(n_0)$, $\hat N_{NS}(n_0)$ are defined in the following way. We chose to normalize the fermion two point correlators as
$\langle \psi(z_2)\psi(z_3)\rangle\rightarrow\frac{z_2}{z_3-z_2}$ as $z_2\rightarrow z_3$. Under this normalization, the number operators are 
\begin{align}
&\hat N_R(n_0)=\oint\oint \frac{dz_2dz_3}{(2\pi i)^2} \frac{1}{(z_3-z_2)}\bigg(\frac{z_2}{z_3}\bigg)^{n_0}\frac{1}{z_3}\psi(z_2)\psi(z_3) \ , \label{eq:defNR} \\
&\hat N_{NS}(n_0)=\oint\oint \frac{dz_2dz_3}{(2\pi i)^2} \frac{1}{(z_3-z_2)}\bigg(\frac{z_2}{z_3}\bigg)^{n_0}\frac{\sqrt{z_2}}{\sqrt{z_3}}\frac{1}{z_3} \psi(z_2)\psi(z_3)\ . \label{eq:defNNS}
\end{align}
Here, $n_0\in Z_{\ge2}$ is a large positive integer, and the integration contour is chosen as $\alpha<|z_2|=1-\epsilon<|z_3|=1+\epsilon<\alpha^{-1}$. The scaling variable is chosen as
\begin{align}
&t=\frac{4\pi r}{L}n_0 \ ,  \  { \rm for}  \ N_{NS}(t,\alpha) \ , \\
&t=\frac{4\pi r}{L}\left(n_0+\frac{1}{2}\right) \ ,   \  { \rm for} \ N_{R}(t,\alpha) \ , 
\end{align}
since the fermion energies at the threshold in the two sectors are respectively  $\frac{2\pi}{L} n_0$ and $\frac{2\pi}{L}(n_0+\frac{1}{2})$. The correlation functions required can be extracted by taking the $\tau=iT \rightarrow +i\infty$ limit on the torus correlators given in~\cite{DiFrancesco:1997nk}. There are some subtitles related to $N_R(t,\alpha)$, since it needs to be extracted by combining the $(0,0)$ and $(0,\frac{1}{2})$ sectors. 

We introduce the two-point function
\begin{align}\label{eq:2ptNS}
f(\alpha)=\langle \Omega_{NS}|\sigma(z_1)\sigma(z_4)|\Omega_{NS} \rangle \ , 
\end{align}
for which we do not need its precise form. Then, one has
\begin{align}  \label{eq:2ptcyl}
\langle \Omega_R^{\pm}|\sigma(z_1)\sigma(z_4)|\Omega_R^{\pm}\rangle=f(\alpha)\alpha^{\pm\frac{1}{2}} \ . 
\end{align}
In terms of the above, one has
\begin{align}
&\frac{\langle \Omega_{NS} |\sigma(z_1)\psi(z_2)\psi(z_3)\sigma(z_4)|\Omega_{NS}\rangle}{f(\alpha)} \nonumber \\ 
&\equiv G_3(z_2,z_3,\alpha)=\frac{\sqrt{z_2}\sqrt{z_3}}{2(z_3-z_2)}\bigg(\sqrt{\frac{z_2-\alpha}{\alpha z_2-1}}\sqrt{\frac{\alpha z_3-1}{z_3-\alpha}}+\sqrt{\frac{z_3-\alpha}{\alpha z_3-1}}\sqrt{\frac{\alpha z_2-1}{z_2-\alpha}}\bigg) \ .
\end{align}
Here, the square roots $\sqrt{z_2}$ and $\sqrt{z_3}$ are defined with the $(0,2\pi)$ branch. Then, when $\alpha<|z_2|<\frac{1}{\alpha}$ and $\alpha<|z_3|<\frac{1}{\alpha}$, the branch singularities all cancel, and one is actually in the R sector. Similarly, one also needs the following 
\begin{align}
&\frac{\langle \Omega_{R}^+ |\sigma(z_1)\psi(z_2)\psi(z_3)\sigma(z_4)|\Omega_{R}^+\rangle\pm\langle \Omega_{R}^- |\sigma(z_1)\psi(z_2)\psi(z_3)\sigma(z_4)|\Omega_{R}^-\rangle}{f(\alpha)} \equiv G_{\pm}(z_2,z_3,\alpha)\nonumber \\ 
&=\frac{1}{2(z_3-z_2)}\bigg((z_2\sqrt{\alpha}\pm z_3\alpha^{-\frac{1}{2}})\sqrt{\frac{z_2-\alpha}{\alpha z_2-1}}\sqrt{\frac{\alpha z_3-1}{z_3-\alpha}}+(z_2 \leftrightarrow z_3)\bigg) \ . \label{eq:4ptR}
\end{align}
In terms of the notations in~\cite{DiFrancesco:1997nk}, one has $G_{+}=(\sqrt{\alpha}+\alpha^{-\frac{1}{2}})G_{2}$   
and $G_{-}=(\sqrt{\alpha}-\alpha^{-\frac{1}{2}})G_{1}$, where $2$ and $1$ denote the $(R,NS)$ and $(R,R)$ sectors on the torus. Due to the square roots, when $\alpha<|z_2|<\frac{1}{\alpha}$ and $\alpha<|z_3|<\frac{1}{\alpha}$, one is in the NS sector.

Given the four-point correlators, we can compute the number observable averages by using the definitions Eq.~(\ref{eq:defNR}), Eq.~(\ref{eq:defNNS}). In the resulting double integrals over the four-point corelators, we deform the contours of $z_2$, $z_3$ to the inside and outside of the unit circle and pick up the branch cuts along $(0,\alpha)$ and $(\alpha^{-1},\infty)$. We simplify further by changing the integration variables along the branch cuts to  $z_2=\alpha \alpha^{t_1}$ and $z_3=\frac{1}{\alpha \alpha^{t_2}}$, with $0<t_1, t_2<\infty$. This way, one ends up at 
\begin{align}
N_{NS}(t,\alpha)=\frac{e^{-t}}{2\pi^2}\int_0^{\infty}&d^2t \frac{e^{-\frac{t t_{12}}{2}}\alpha^{3+\frac{3t_{12}}{2}}\ln^2 \alpha}{(1-\alpha^{t_{12}+2})^2}\nonumber \\ \times&\bigg(\sqrt{\frac{\alpha^2(1-\alpha^{t_1})(1-\alpha^{t_2})}{(1-\alpha^{2+t_1})(1-\alpha^{2+t_2})}}+\sqrt{\frac{(1-\alpha^{2+t_1})(1-\alpha^{2+t_2})}{\alpha^2(1-\alpha^{t_1})(1-\alpha^{t_2})}}\bigg) \ . 
\end{align}
Here, $t_{12}=t_1+t_2$ and $d^2t=dt_1dt_2$. It is very similar to the two-particle form-factor integrals in the Ising model. In the scaling limit $\alpha \rightarrow 1$, by expanding the integrands one obtains
\begin{align}\label{eq:cylinNS}
N_{NS}(t,\alpha)=\mathfrak{f}_{+}(t)+(1-\alpha)\partial_t\mathfrak{f}_+(t)+{\cal O}((1-\alpha)^2) \ . 
\end{align}
The parameterization $\alpha^{t}$ simplifies the integrands for the $1-\alpha$ corrections, since
\begin{align}
\sqrt{\frac{\alpha(1-\alpha^t)}{1-\alpha^{2+t}}} \rightarrow \sqrt{\frac{t}{t+2}}-\frac{(1-\alpha)^2}{12} (t+1) \sqrt{\frac{t}{t+2}}+{\cal O}((1-\alpha)^3) \ . 
\end{align}
The above in fact establishes a simpler representation of the scaling function $\mathfrak{f}_+(t)$
\begin{align}
&\mathfrak{f}_+(t)\equiv \frac{e^{-t}}{2\pi^2}\int_0^{\infty}\int_0^{\infty} \frac{e^{-t_{12}t/2}}{(t_{12}+2)^2} dt_1dt_2 \bigg(\sqrt{\frac{t_1}{t_1+2}}\sqrt{\frac{t_2}{t_2+2}}+\sqrt{\frac{t_1+2}{t_1}}\sqrt{\frac{t_2+2}{t_2}}\bigg)  \ ,  \\ 
&\equiv \frac{1}{2\pi^2}\int_0^{\infty}\int_0^{\infty}\frac{dx_1dx_2}{(x_1+x_2)^2}e^{-\frac{t}{4}\left(x_1+\frac{1}{x_1}+x_2+\frac{1}{x_2}\right)}  \ , \\
&\equiv\frac{t}{2\pi^2}K_0\left(\frac{t}{2}\right)K_1\left(\frac{t}{2}\right)-\frac{t^2}{4\pi^2}\left(K_1^2\left(\frac{t}{2}\right)-K_0^2\left(\frac{t}{2}\right)\right) \ . 
\end{align}
To reach the second integral, one changes the variables as $t_i \rightarrow \frac{1}{2}(x_i+x_i^{-1})-1$. The equivalence to the Barnes representations in Eq.~(\ref{eq:scalingsimple})  manifests after Mellin-transforming the scaling variable $t$. 

Strikingly, the scaling function $\mathfrak{f}_+(t)$, a quantity defined in the Ising CFT, is in fact identical to the two-particle form-factor integrals in the massive-fermion Ising QFT. For example, the first integral above is just the $\frac{1}{2\pi^2}\psi_2$, where the functions $\psi_n$ are defined in Eq.~(4.7) of~\cite{McCoy:1976cd}.
The second integral is just the two particle contribution to the potential function $\phi_{\pm}= {\rm Tr} \ln (1\pm \lambda K)$, where $K$ is defined in Eq.~(\ref{eq:defKgeneral}) with $E=1$.
Its simplification to the Bessel functions is well known~\cite{Wu:1975mw}. Using the equations Eq.~(\ref{eq:fminusdif}), Eq.~(\ref{eq:QLeading2}), one can simplify the  $\mathfrak{f}_-(t)$ and $\mathfrak{g}(t)$ as well
\begin{align}
\mathfrak{f}_-(t)=\mathfrak{f}_+(t)-\frac{1}{\pi^2}K_0^2\left(\frac{t}{2}\right) \ , \\
\mathfrak{g}(t)=-\frac{1}{\pi^2}\int_t^{\infty}\frac{K_0^2\left(\frac{v}{2}\right)}{v}dv \ . 
\end{align}
Clearly, the $\mathfrak{g}(t)$ is more transcendental than $\mathfrak{f}_{\pm}(t)$. On the other hand, both of the $\mathfrak{f}_{\pm}(t)$ can be expressed in terms of the Bessel-$K$ functions. This suggests that the $\mathfrak{f}_{-}(t)$ is also related to the number observables on the cylinder, which we show now.

To obtain $\mathfrak{f}_{-}(t)$, one needs the R sector. It is convenient to define
\begin{align} \label{eq:defNpmcy}
N_{\pm}(t,\alpha)=\frac{\langle \Omega_{R}^+ |\sigma(z_1)\hat N_{NS}(n_0)\sigma(z_4)|\Omega_{R}^+\rangle\pm\langle \Omega_{R}^- |\sigma(z_1)\hat N_{NS}(n_0)\sigma(z_4)|\Omega_{R}^-\rangle}{f(\alpha)}  \ .
\end{align}
Using the four-point correlators Eq.~(\ref{eq:4ptR}) and deform the integration contours as before, one has
\begin{align}
&N_{\pm}(t,\alpha)=\frac{e^{-t}\ln^2 \alpha}{2\pi^2} \int_{0}^{\infty} d^2t e^{-\frac{t_{12}t}{2}}\alpha^{2+t_{12}}\frac{\alpha^{\frac{5}{2}+t_{12}}\pm\sqrt{\alpha^{-1}}}{(1-\alpha^{2+t_{12}})^2} \sqrt{\frac{\alpha(1-\alpha^{t_1})}{1-\alpha^{2+t_1} }}\sqrt{\frac{\alpha(1-\alpha^{t_2})}{1-\alpha^{2+t_2} }}\nonumber \\ 
&+\frac{e^{-t}\ln^2\alpha}{2\pi^2} \int_{0}^{\infty} d^2t e^{-\frac{t_{12}t}{2}} \alpha^{2+t_{12}}\frac{\sqrt{\alpha}\pm\alpha^{\frac{3}{2}+t_{12}}}{(1-\alpha^{2+t_{12}})^2} \sqrt{\frac{1-\alpha^{2+t_1}}{\alpha(1-\alpha^{t_1})}}\sqrt{\frac{1-\alpha^{2+t_2}}{\alpha(1-\alpha^{t_2})}} \ .
\end{align}
By expanding, one again has
\begin{align} \label{eq:asymNp}
N_+(t,\alpha)=2\mathfrak{f}_+(t)+2(1-\alpha)\partial_t\mathfrak{f}_+(t)+{\cal O}((1-\alpha)^2) \ . 
\end{align}
On the other hand, from $N_-(t,\alpha)$, one sees the appearance of $\mathfrak{f}_-(t)$: 
\begin{align}\label{eq:asymNm}
N_-(t,\alpha)=(\alpha-1)\mathfrak{f}_-(t)+{\cal O}((1-\alpha)^2) \ . 
\end{align}
For the above, one needs the following integral representation
\begin{align}
&\mathfrak{f}_-(t)-\mathfrak{f}_+(t) \nonumber \\ 
&=\frac{1}{2\pi^2}\int_1^{\infty}\int_1^{\infty} \frac{e^{-(t_1+t_2)t/2}}{t_1+t_2} dt_1dt_2 \bigg(\sqrt{\frac{t_1-1}{t_1+1}}\sqrt{\frac{t_2-1}{t_2+1}}-\sqrt{\frac{t_1+1}{t_1-1}}\sqrt{\frac{t_2+1}{t_2-1}}\bigg) \ .
\end{align}
There seems to be a $1/t$ singularity at small-$t$, but that was canceled among the two terms, with an increased logarithmic singularity. After introducing $2t_i=x_i+x_i^{-1}$, one has
\begin{align}
\mathfrak{f}_-(t)-\mathfrak{f}_+(t)=-\frac{1}{4\pi^2}\int_0^{\infty}\int_0^{\infty}\frac{dx_1dx_2}{x_1x_2}e^{-\frac{t}{4}\left(x_1+\frac{1}{x_1}+x_2+\frac{1}{x_2}\right)}=-\frac{1}{\pi^2}K_0^2\left(\frac{t}{2}\right) \ , 
\end{align}
as expected. Given the above, one can write
\begin{align}
&N_R(t,\alpha)=\frac{N_+(t,\alpha)-N_-(t,\alpha)}{2\alpha^{-\frac{1}{2}}}\nonumber \\ 
&=\mathfrak{f}_+(t) +(1-\alpha)\left(\frac{d\mathfrak{f}_+}{dt}+\frac{\mathfrak{f}_--\mathfrak{f}_+}{2}\right)(t)+{\cal O}((1-\alpha)^2) \ . \label{eq:cylinR}
\end{align}
Equations Eq.~(\ref{eq:cylinNS}) and Eq.~(\ref{eq:cylinR}) are the major results of this appendix. 

In fact, there is another representation of $\mathfrak{f}_-(t)$ that is more similar to the massive theory. Using the relations for the two-point correlator Eq.~(\ref{eq:2ptcyl}), one can express $\mathfrak{f}_-(t)$ as 
\begin{align} \label{eq:fminusratio}
\mathfrak{f}_-(t)=\lim_{\alpha\rightarrow 1}\frac{\langle \Omega_{NS}|\sigma(z_1)\hat N_R(n_0)\sigma(z_4)|\Omega_{NS} \rangle-\langle \Omega_{R}^-|\sigma(z_1)\hat N_{NS}(n_0)\sigma(z_4)|\Omega^-_{R} \rangle}{\langle \Omega_{NS}|\sigma(z_1) \sigma(z_4)|\Omega_{NS}\rangle-\langle \Omega_{R}^-|\sigma(z_1) \sigma(z_4)|\Omega_{R}^-\rangle} \ . 
\end{align}
Indeed, using the definitions 
Eq.~(\ref{eq:defNpmcy}), the relations Eq.~(\ref{eq:2ptNS}), Eq.~(\ref{eq:2ptcyl}) and the asymptotics Eq.~(\ref{eq:cylinNS}), Eq.~(\ref{eq:asymNp}), Eq.~(\ref{eq:asymNm}), 
one has 
\begin{align}
&\frac{\langle \Omega_{NS}|\sigma(z_1)\hat N_R(n_0)\sigma(z_4)|\Omega_{NS} \rangle-\langle \Omega_{R}^-|\sigma(z_1)\hat N_{NS}(n_0)\sigma(z_4)|\Omega^-_{R} \rangle}{\langle \Omega_{NS}|\sigma(z_1) \sigma(z_4)|\Omega_{NS}\rangle-\langle \Omega_{R}^-|\sigma(z_1) \sigma(z_4)|\Omega_{R}^-\rangle} \nonumber \\ 
&=\frac{N_{NS}(t,\alpha)-\frac{N_+(t,\alpha)-N_-(t,\alpha))}{2}}{1-\frac{1}{\sqrt{\alpha}}} \rightarrow \frac{N_-(t,\alpha)+{\cal O}((1-\alpha)^2))}{2(1-\alpha^{-\frac{1}{2}})} \rightarrow \mathfrak{f}_-(t) \ . 
\end{align}
Notice that Eq.~(\ref{eq:fminusratio}) is very similar to the following relation in the massive theory: 
\begin{align}
\lim_{r\rightarrow0} \frac{F_+(r)N_+(t,r)-F_-(r)N_-(t,r)}{F_+(r)-F_-(r)}=\mathfrak{f}_-(t) \ .
\end{align}
To show this, one needs the Eq.~(\ref{eq:finalasymptotics}) and the well known asymptotics of the two-point correlators
\begin{align}
F_{\pm}(r) \propto\frac{1}{r^{\frac{1}{4}}}\bigg(1\pm\frac{r}{2}\Omega+{\cal O}(r^2)\bigg) \ . 
\end{align}
From the above, the similarity to the Eq.~(\ref{eq:fminusratio}) becomes manifest.

\section{Resumming the collapsed power corrections} \label{sec:expansiondetail}
This appendix explains the resummation procedure of the power corrections that collapse to their destination locations as $\sigma \rightarrow 1$. We first consider the profile function $\varphi(r)$ which satisfies the Sinh Gordon equation
\begin{align}
\left(\frac{d^2}{dr^2}+\frac{1}{r}\frac{d}{dr}\right)\varphi(r)=\frac{1}{2}\sinh 2\varphi(r) \ . 
\end{align}
To start, we write $\eta=e^{-\varphi}=Br^{\sigma}\eta_0(1+\eta_1+....)$ as an ansartz. Here, $\eta_0=1+{o}(1)$ when $0<\sigma<1$, but it remains ${\cal O}(1)$ as $\sigma \rightarrow 1$.  And $\eta_1$ is the leading power corrections that do not collapse to ${\cal O}(1)$ as $\sigma \rightarrow1$. The purpose is to solve for $\eta_0$ and $\eta_1$.  Let's expand the $\sinh$ function to the linear order in $\eta_1$. This leads to
\begin{align}
&\frac{1}{2}\sinh 2\phi=\frac{1}{4B^2r^{2\sigma}\eta_0^2}-\frac{\eta_1}{2B^2 r^{2\sigma}\eta_0^2}-\frac{B^2r^{2\sigma}}{4}\eta_0^2-\frac{B^2\eta_0^2r^{2\sigma}}{2}\eta_1+.... \ . \label{eq:expandsinh}
\end{align}
On the other hand, for the potential function one has 
\begin{align}
\phi=-\ln \eta=-\ln B-\sigma \ln r-\ln \eta_0-\eta_1+.... \ . 
\end{align}
Now, the $-\ln B$ and $-\sigma \ln r$ are killed by the Laplacian, so we start from the $-\ln \eta_0$ term. Our assumption is, as $\sigma \rightarrow 1$, the $\ln \eta_0$ collapse to ${\cal O}(1)$. As such, applying the Laplacian, the power decreases to $-2$. 
On the right hand side of Eq.~(\ref{eq:expandsinh}), only the $r^{-2\sigma}\eta_0^{-2}$ term matches this $-2$, since $\eta_1$ is assumed to be a power correction. As such, for the $\eta_0$, one obtains the following non-linear equation
\begin{align}
&\left(\frac{d^2}{dr^2}+\frac{1}{r}\frac{d}{dr}\right)\ln \eta_0=-\frac{1}{4B^2r^{2\sigma}\eta_0^2} \ , \\
&\eta_0(r)|_{r \rightarrow 0}=1+{o}(r^{2-2\sigma}) \ . 
\end{align}
The ${o}(r^{2-2\sigma)}$ naturally follows from the differential equation.  The solution can be found by changing the variable to $r^{2-2\sigma}$
\begin{align}
\eta_0(r)=1+br^{2-2\sigma} \ , \  b=-\frac{1}{16B^2(1-\sigma)^2} \ . 
\end{align}
Given the above, we proceed to $\eta_1$. The $r^{2\sigma}\eta_0^2$ term in Eq.~(\ref{eq:expandsinh}) will source the ${\cal O}(r^4)$ corrections by power counting. As such, it is natural to assume that $\eta_1$ is also at this order and satisfies the linear equation
\begin{align}
\left(\frac{d^2}{dr^2}+\frac{1}{r}\frac{d}{dr}\right)\eta_1(r)=\frac{\eta_1(r)}{2B^2r^{2\sigma}(1+br^{2-2\sigma})^2}+\frac{B^2 r^{2\sigma}(1+b r^{2-2\sigma})^2}{4} \ , \label{eq:diffeeta5}\\  
\eta_1(r)|_{r\rightarrow 0}={\cal O}(r^{2+2\sigma}) \ .
\end{align}
On the other hand, if there were power corrections before the ${\cal O}(r^4)$ but after $\eta_0$, then there would be no source for them, and the first of them must satisfy the homogeneous equation
\begin{align}
\left(\frac{d^2}{dr^2}+\frac{1}{r}\frac{d}{dr}\right)f(r)=\frac{f(r)}{2B^2r^{2\sigma}(1+br^{2-2\sigma})^2} \ . 
\end{align}
It is exactly of the type Eq.~(\ref{eq:diffgen}) with $\alpha=0$. But it has only two solutions
that are ${\cal O}(1)$ as $\sigma \rightarrow 1$. As such, 
they can not represent power corrections between ${\cal O}(1)$ and ${\cal O}(r^4)$, as all the ${\cal O}(1)$ corrections are already resummed by the $\eta_0$. As such, we conclude that the first correction $\eta_1$ on top of $\eta_0$ is indeed at ${\cal O}(r^4)$ and satisfies Eq.~(\ref{eq:diffeeta5}). The $\eta_1r^{2\sigma}$ term in Eq.~(\ref{eq:expandsinh}) will then source the ${\cal O}(r^8)$ corrections, which is the location of the next correction $\eta_2$ after $\eta_1$.   By changing the coordinate to $w=-br^{2-2\sigma}$, it is ready to see that the solution is 
\begin{align}
\eta_1(r)=\frac{B^2r^{2\sigma+2}}{16(1+\sigma)^2}\frac{1-(1+2\sigma)w+\frac{(\sigma+1)^2(5-2\sigma)}{(\sigma-3)^2}w^2-\frac{(\sigma+1)^2}{(\sigma-3)^2}w^3}{1-w} \ , 
\end{align}
which leads to the $\eta^{(5)}(r)$ in the Eq.~(\ref{eq:etafourth}) after multiplying back the $Br^{\sigma}\eta_0$. This procedure can be continued to an arbitrary power. Notice that for the $B(\sigma)$ given in Eq.~(\ref{eq:defB}), all the singularities at $\sigma=1$ cancel in $B\eta_0$ and $B\eta_0\eta_1$.

For the two-variable equations, this procedure generalizes as well. Here we show the arguments for $V^{(0)}(r,Y)$. The point is that, the source terms due to subleading corrections in $\cosh 2\varphi$ start to contribute only from the quartic power and higher. As such, $V^{(0)}(r,Y)$ and $V^{(2)}(r,Y)$ must satisfy the homogeneous equations
\begin{align}
\left(\frac{\partial^2}{\partial r^2}+\frac{1}{r}\frac{\partial}{\partial r}-\frac{1}{r^2}\frac{\partial^2}{\partial Y^2}\right)V^{(i)}(r,Y)=\frac{1}{2B^2r^{2\sigma}(1+br^{2-2\sigma})^2}V^{(i)}(r,Y) \ . \label{eq:differVappen}
\end{align}
The crucial information here are in the scaling function $f_+(t,\sigma)-f_-(t,\sigma)$. At the $t^{1-\sigma}$ order, there is a contribution give by the last term of Eq.~(\ref{eq:f0minus}), which we denote as $c_0 r^{1-\sigma}e^{(1-\sigma)Y}$. This term must be the lowest contribution in $V^{(0)}(r,Y)$ when expanded in $r^{1-\sigma}$. If not, then we expand the $V^{(0)}(r,Y)$ to the lowest order in $r^{1-\sigma}$, say at $r^{(1-\sigma)\alpha}$. If $\alpha>0$, it must be of the form $\sinh (1-\sigma)\alpha Y \times  r^{(1-\sigma)\alpha}$. But such a term would have appeared in the scaling function already, which is a contradiction. On the other hand, if $\alpha=0$, then in the scaling function, there are indeed contributions given by the first terms in Eq.~(\ref{eq:f0plus}) and Eq.~(\ref{eq:f0minus}), but they were already separated out in the decomposition ${\cal Q}(r,Y)=\frac{\lambda}{2}\partial_\lambda \varphi(r)+\frac{\lambda}{2}\partial_{\lambda} \varphi\bigg|_{\ln r} Y+V(r,Y)$, see Eq.~(\ref{eq:decomY}) and Eq.~(\ref{eq:decompPsi}).  The constant and $\ln t$ terms in the scaling functions,  when propagated to higher orders, are completely captured by the combination $\frac{\lambda}{2}\partial_\lambda \varphi(r)+\frac{\lambda}{2}\partial_{\lambda} \varphi \bigg|_{\ln r} Y$.  $V(r,Y)$ only sees the scaling functions at $t^{1-\sigma}$ and higher.  Now, assume that
\begin{align}
V^{(0)}(r,Y)=2c_0r^{1-\sigma}\sinh (1-\sigma) Y+ r^{1-\sigma}\sum_{n=1}^{\infty}r^{n(1-\sigma)}v_n(Y)  \ , 
\end{align}
plug it into the equation Eq.~(\ref{eq:differVappen}), by matching the orders in $r^{1-\sigma}$ we found that 
\begin{align}
((n+1)^2(1-\sigma)^2-\partial_Y^2)v_n(Y)= q_n(c_0\sinh (1-\sigma)Y , v_1,...v_{n-1}) \ , 
\end{align}
where $q_n$ is a linear function in its arguments. Now, we claim that $v_n(Y)$ depends on $Y$ only through $\sinh(1-\sigma)Y$. Indeed, if not, then at the smallest $n$ where this happens, it must due to the homogeneous solution $\sinh (n+1)(1-\sigma)Y$ for $v_n$. But then this term would appeared in the scaling function again, which leads to a contradiction. As such, we can write
\begin{align}
V^{(0)}(r,Y)=V_0(r)\sinh (1-\sigma)Y  \ .
\end{align}
Plug into Eq.~(\ref{eq:differVappen}), the $V_0(r)$ satisfies 
\begin{align}
\bigg(\frac{d^2}{dr^2}+\frac{1}{r}\frac{d}{dr}-\frac{(1-\sigma)^2}{r^2}\bigg)V_0(r)=\frac{1}{2B^2r^{2\sigma}(1+br^{2-2\sigma})^2}V_0(r) \ , \\
V_0(r)|_{r\rightarrow 0}=2c_0r^{1-\sigma} \ . 
\end{align}
This equation is of the type Eq.~(\ref{eq:diffgen}) and is solved in Eq.~(\ref{eq:solutionV0}), if we recall from the Eq.~(\ref{eq:f0minus}) the following value
\begin{align}
2c_0=\frac{2^{3 \sigma-3}  \sin \left(\frac{\pi  \sigma}{2}\right) \Gamma \left(\frac{\sigma-1}{2}\right)^2}{\pi ^2} \ .
\end{align}
The solution to $V^{(2)}(r,Y)$ follows essentially the same procedure. 

\section{Frequently used linear equations} \label{sec:diffalpha}

In this work, the following differential equations will be frequently used 
\begin{align}
&\bigg(\frac{d^2}{dr^2}+\frac{1}{r}\frac{d}{dr}-\frac{\alpha^2}{r^2}\bigg){\cal F}_{\alpha}(r)=\frac{{\cal F}_{\alpha}(r)}{2B^2r^{2\sigma}(1+br^{2-2\sigma})^2} \ , \label{eq:diffgen} \\ 
&b=-\frac{1}{16B^2(1-\sigma)^2} \ . 
\end{align}
For $\sigma<1$, the small-$r$ singularity continues to be regular. By writing ${\cal F}_{\alpha}(r)=r^{\alpha}F_{\alpha}(r)$ and change the variable to $w=-br^{2-2\sigma}$, the equation can be simplified to 
\begin{align}
wF_{ww}+aF_{w}=\frac{2}{(1-w)^2}F  \ ,  \ a=1+\frac{\alpha}{1-\sigma} \ . 
\end{align}
The two linearly independent solutions are given by
\begin{align}
F_r(w)=\frac{1+w}{1-w} \  , \ F_s(w)= F_r(w)\ln w +\frac{4}{1-w} \ ,
\end{align}
for $a=1$, and 
\begin{align}
F_r(w)=1+\frac{2w}{1-w}\frac{1-\sigma}{1-\sigma+\alpha}\  , \ F_s(w)= \frac{w^{1-a} (a (1-w)-2)}{w-1} \ ,
\end{align}
for $a>1$.

\section{Numerical evidences for the asymptotic formula} \label{sec:numer}
\begin{figure}[htbp]
    \centering
    \includegraphics[height=5.0cm]{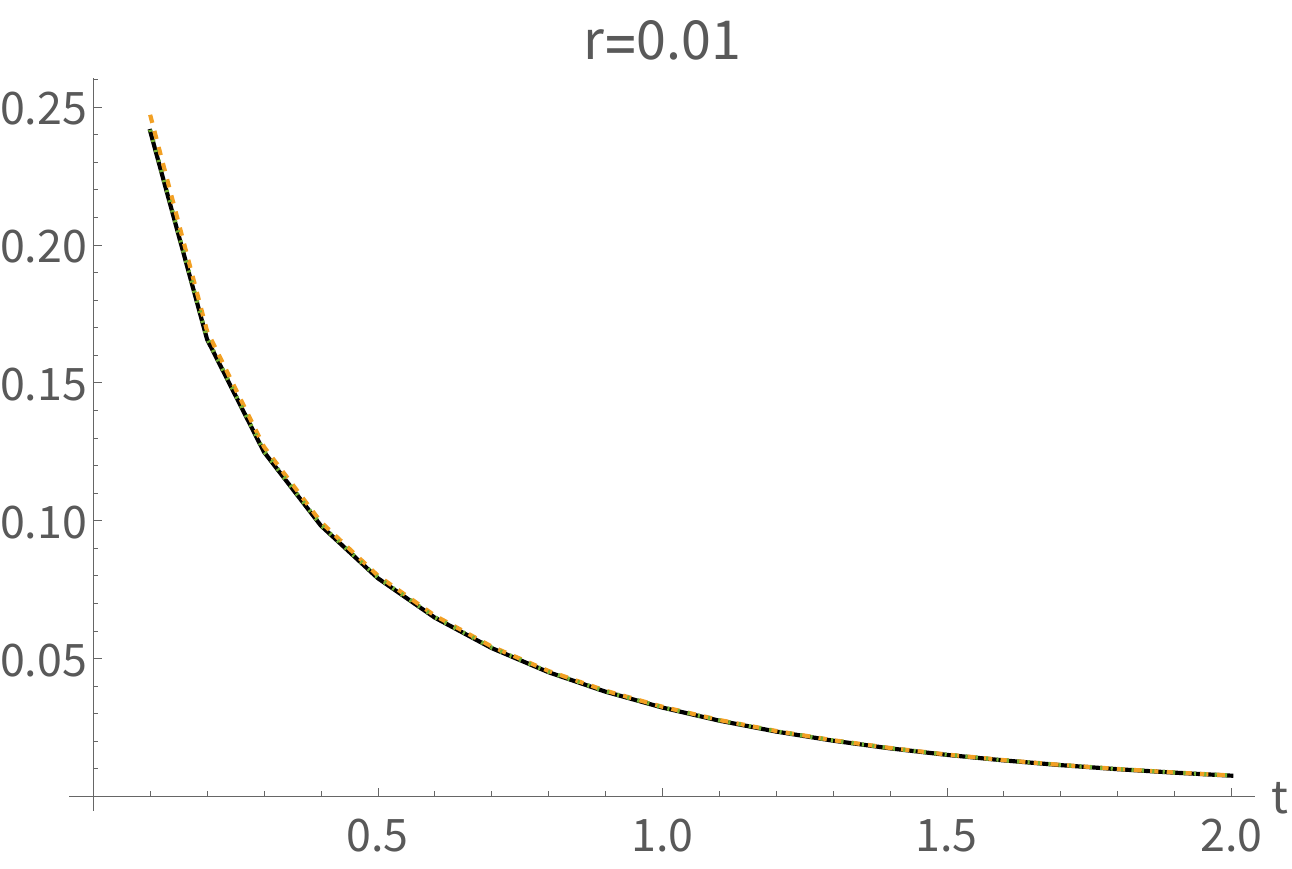}
    \includegraphics[height=5.0cm]{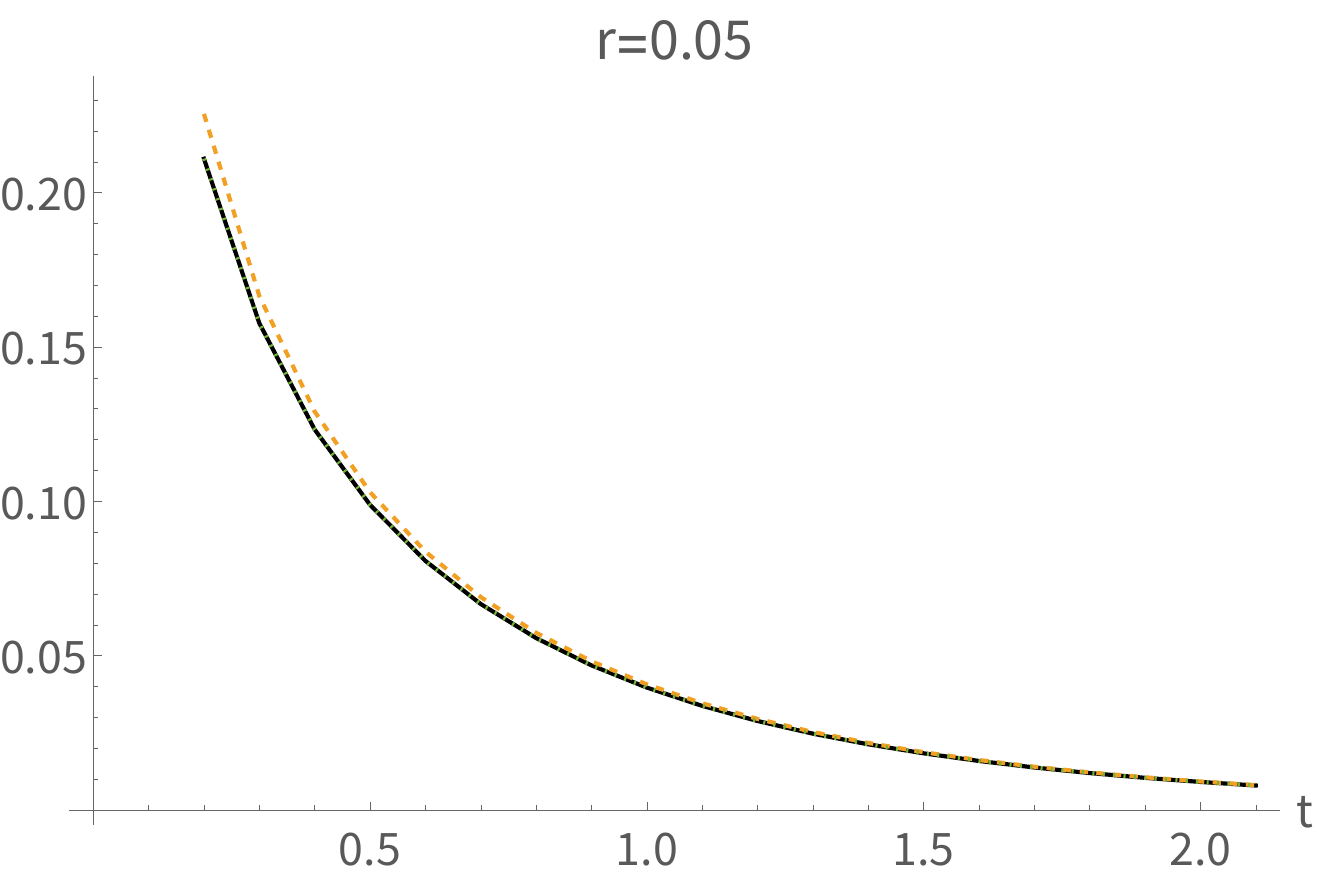}
    \includegraphics[height=5.0cm]{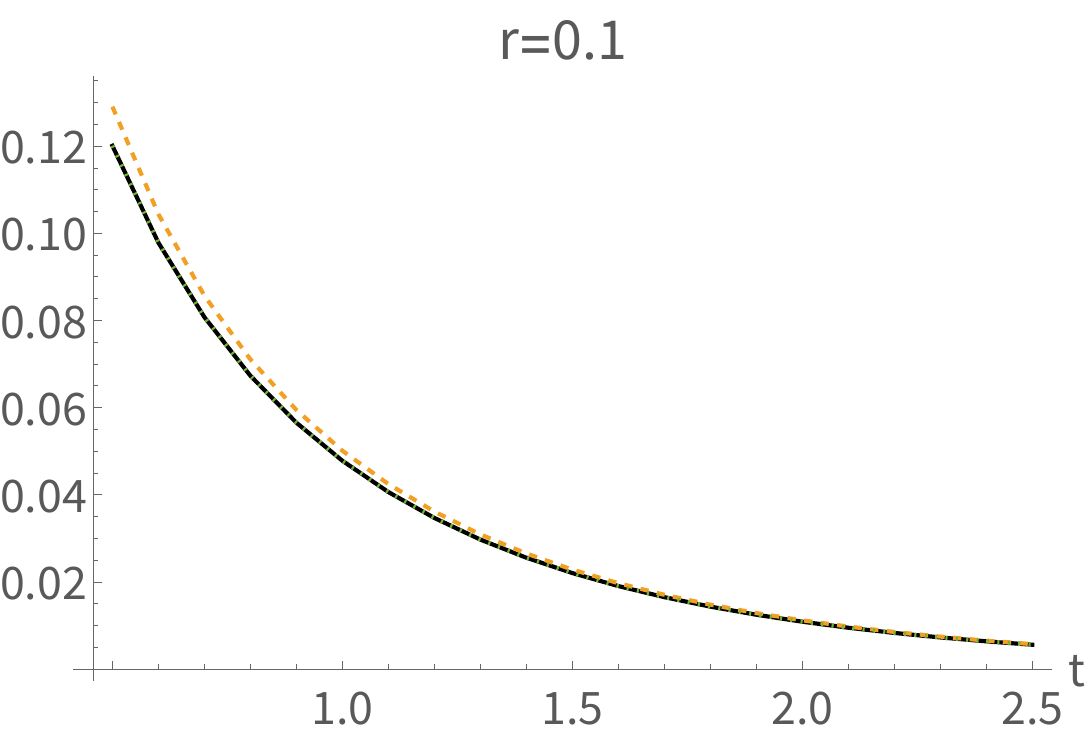}
    \caption{The comparison at $r=0.01$ (upper left), $r=0.05$ (upper right) and $r=0.1$ (lower), between the $N_{3}(t,r)$ (black solid), $N_{\rm asym}(t,r)$ (dotted) and the  ``wrong'' asymptotic formula with $\mathfrak{g}(t)$ removed (orange-dashed). The horizontal-axis is the scaling variable $t$. Each of the figure are composed of $20$ or $21$ data points at $t \in 0.1\mathbb{Z}$ connected by straight lines.  
    The ranges of $t$ are: $0.1\le t\le2$ ($r=0.01$), $0.2\le t\le 2.1$ ($r=0.05$) and $0.5\le t \le 2.5$ ($r=0.1$). The lower values of $t$ are to guarantee that $2t/r =e^Y \ge 8$. In all these plots, the three particle truncation $N_3(t,r)$ and the asymptotic formula $N_{\rm asym}(t,r)$ are sufficiently close to each other and hard to distinguish. At $r=0.05$ and $r=0.1$, the discrepancies are below $5\times 10^{-4}$, while at $r=0.01$ the discrepancies increase to $5\times 10^{-3}$. On the other hand, the plots with $\mathfrak{g}(t)$ removed deviate from the other two curves as $t$ decrease, and the discrepancies are quite significant at $r=0.05$ and $r=0.1$. }
    \label{fig:r1}
\end{figure}

To convince the readers that the Eq.~(\ref{eq:finalasymptotics}) is the correct asymptotic formula for the number observables, here we compare it numerically with the form factor expansion truncated at the three particle level for $N_-$ (above critical temperature). Usually, $r\ge 0.01$, the form factor expansion in integrable QFTs truncated at the three particle level works very good for $r\ge 0.01$, see Ref.~\cite{Zamolodchikov:1990bk} for a good example. We have checked that for the original two point correlator, at $r=0.01$, the three particle truncation agrees with the ${\cal O}(r^8)$-order short distance asymptotics given in Ref.~\cite{Au-Yang2002}, with a mismatch up to $5\times 10^{-4}$, and at $r=0.05$ the mismatch decreases to about $10^{-5}$. We denote the right hand side of the Eq.~(\ref{eq:finalasymptotics})
as 
\begin{align}
N_{\rm asym}(t,r)=\mathfrak{f}_{+}(t)+\frac{r}{2-r\Omega} \bigg( \left(\mathfrak{f}_+(t)-\mathfrak{f}_-(t)\right)\Omega+\mathfrak{g}(t)\bigg) \ .
\end{align}
 The three-particle truncation is defined as
\begin{align}
&N_3(t,r) \nonumber \\ 
&=\frac{\frac{1}{2\pi}\int_{\frac{2t}{r}}^{\infty}\frac{dx}{x} e^{-\frac{r}{2}(x+x^{-1})}+ \frac{3}{3!(2\pi)^3}\int_{\frac{2t}{r}}^{\infty} \frac{dx_1}{x_1}\int_0^{\infty}\frac{dx_2dx_3}{x_2x_3}\prod_{i<j} \left(\frac{x_i-x_j}{x_i+x_j}\right)^2 e^{-\frac{r}{2}\sum_{i}^3(x_i+x_i^{-1})}}{\frac{1}{2\pi}\int_0^{\infty}\frac{dx}{x} e^{-\frac{r}{2}(x+x^{-1})}+ \frac{1}{3!(2\pi)^3}\int_0^{\infty}\frac{dx_1dx_2dx_3}{x_1x_2x_3}\prod_{i<j}\left(\frac{x_i-x_j}{x_i+x_j}\right)^2 e^{-\frac{r}{2}\sum_{i}^3(x_i+x_i^{-1})} } \ . 
\end{align}
As expected, the agreement with the asymptotic formula is rather good.

For example, at $r=0.1$ and $t=0.5$, one has $N_{\rm asym}(0.5,0.1)=0.12005$ while $N_{3}(0.5,0.1)=0.120036$. At $r=0.1$ and $t=2$ one has $N_{\rm asym}(2,0.1)=0.0108317$, while the $N_3(2,0.1)=0.010831$. At larger $r$, the $N_{\rm asym}(t,r)$ is still not very bad. For example, at $r=1$ and $t=0.5$, one has $N_3(0.5,1)=0.500037$. This is expected, as $t=\frac{r}{2}$ means $Y=0$, one simply measures half of the mean particle number, which is almost $0.5$ in the large distance region dominated by the one-particle term. At this value, the asymptotic formula gives $N_{\rm asym}(0.5,1)=0.507517$, which is still not very bad. In the Fig.~\ref{fig:r1}, we provide three plots at $r=0.01$, $r=0.05$, $r=0.1$ that depict the $N_3(t,r)$ and $N_{\rm asym}(t,r)$ as functions of $t$.  To show the effect of the function $\mathfrak{g}(t)$, we also include the plots (shown in dashed orange line) for the asymptotic formula with $\mathfrak{g}(t)$ removed. 

It needs to be mentioned that we also checked the asymptotic formula Eq.~(\ref{eq:expandNsmallr}) at fixed $Y$ against the form factor expansion (high-temperature case). As expected, when $Y$ is small, it works better than the fixed $t$ expansion. For example, at $r=0.1$ and $Y=0$ ($t=0.05$), one has $N_{\rm asym}(0.05,0.1)=0.512318$ , while the fixed $Y$ formula Eq.~(\ref{eq:expandNsmallr}) gives $0.511412$. The three particle truncation gives $N_3(0.05,0.1)=0.511434$ and agrees better with the fixed-$Y$ formula. As $Y$ becomes negative, the $N_{\rm asym}$ becomes worse, while the fixed-$Y$ formula is still good, as far as $|Y|$ is not too large.

\bibliographystyle{apsrev4-1} 
\bibliography{bibliography}

\end{document}